\documentclass[a4paper,11pt]{article}
\usepackage{graphicx}
\usepackage{algorithm}
\usepackage{algpseudocode}
\usepackage{amsfonts}
\usepackage{amsmath, amsthm, amssymb}
\usepackage{dsfont}
\usepackage[utf8x]{inputenc}
\usepackage{color}
\usepackage{geometry}
\usepackage{subfigure}
\usepackage{ifthen}
\title{Ideomotor feedback control in a recurrent neural network}
 \author{Mathieu Galtier\footnote{Minds, Jacobs University Bremen / NeuroMathComp, Inria Sophia / UNIC, CNRS,  mathieu.galtier@gmx.com}}

\newcounter{ALC@tempcntr}

\def \eps {\epsilon}

\def \R {\mathbb{R}}

\def \P {\mathbf{P}}

\def \v {\mathbf{v}}
\def \vr {\mathbf{v}^\text{r}}
\def \vp {\mathbf{v}^\text{p}}
\def \va {\mathbf{v}^\text{a}}
\def \vf {\mathbf{v}^\text{f}}

\def \np {n_\text{p}}
\def \nr {n_\text{r}}
\def \na {n_\text{a}}
\def \mup {\mu_\text{p}}
\def \muf {\mu_\text{f}}
\def \mua {\mu_\text{a}}
\def \lambdap {\lambda_\text{p}}
\def \lambdaf {\lambda_\text{f}}
\def \lambdaa {\lambda_\text{a}}

\def \u {\mathbf{u}}

\def \x {\mathbf{x}}

\def \b {\mathbf{b}}
\def \c {\mathbf{c}}
\def \y {\mathbf{y}}
\def \z {\mathbf{z}}

\def \W {\mathbf{W}}

\def \Wrp {{\mathbf{W}^{\text{rp}}}}
\def \Wrr {{\mathbf{W}^{\text{rr}}}}
\def \Wra {{\mathbf{W}^{\text{ra}}}}

\def \Wp {\mathbf{W}^{\text{pr}}}
\def \Wf {\mathbf{W}^{\text{fr}}}
\def \Wprestr {\mathbf{W}_{\shortmid}^{\text{p}}}
\def \Wr {\mathbf{W}^{\text{r}}}
\def \Wa {\mathbf{W}^{\text{ar}}}

\begin{document}
\maketitle
 \begin{abstract}
	The architecture of a neural network controlling an unknown environment is presented. It is based on a randomly  connected recurrent neural network from which both perception and action are simultaneously read and fed back. There are two concurrent learning rules implementing a sort of ideomotor control: (i) perception is learned along the principle that the network should predict reliably its incoming stimuli; (ii) action is learned along the principle that the prediction of the network should match a target time series. The coherent behavior of the neural network in its environment is a consequence of the interaction between the two principles. Numerical simulations show a promising performance of the approach, which can be turned into a local, and better "biologically plausible", algorithm.
 \end{abstract}

\section{Introduction}
Animal life is characterized by the emergence of robust control mechanisms used in a large variety of environments. Beyond evolution which selects the life forms which make the most of their environment, it seems that some animals have the ability to quickly learn to interact constructively with new environments. Most probably, it means that the nervous systems of such animals can perform a sort of blind control of their environments: control is achieved without previous knowledge of the environment. It would surely be of great help to uncover such a mechanism, not only from a biological viewpoint, but also to replicate it for engineering tasks.

Control theory has been a vivid field of research for decades which has provided many useful applications. However, although linear systems are well understood \cite{kwakernaak1972linear, fortmann1977introduction}, non-linear systems still require a significant effort to be dealt with \cite{slotine1991applied, skogestad2007multivariable}. In particular, the control system is often assumed to have some explicit knowledge about the system to be controlled. A recent axis of research focuses on design of control schemes when there is a lot a uncertainties about the system to be controled \cite{aastrom2012introduction}. When nothing is known about the system to be controlled, the most widespread method is to use a Proportional-Integral-Derivative controller \cite{aastrom2006advanced}, although some sophisticated methods have been proposed, see \cite{zhong2006model} for a review. Indeed, computing the best inverse model \cite{jordan1996computational} from an unknown environment is a very difficult task which has received no universal answer so far. 

Neural networks are bio-inspired mathematical object which have good learning capacities \cite{bishop1995neural}. A large number of control algorithms have used their adaptability to model some aspect of the environment and / or to design adaptive controllers. The challenge is to be able to internally predict the outcomes of a potential movement so that the best motor commands can be chosen \cite{kawato1987hierarchical, ge2008adaptive, yang2008output}. Typically, neural networks are used for two purposes: identification of the environment and the actual control which can be computed once a good estimate of the environment has been designed \cite{narendra1990identification,ge2010stable}.

A main drawback of feedforward neural networks is their inability to take time dependencies into account. Although the usual work-around is to use tapped delay lines, a more natural approach would be to use recurrent neural networks (RNN) which are more suited to dynamical systems approximation. Based on the two steps of identification and control, various RNN have been proposed as controllers e.g. \cite{chow1998recurrent, wang2006learning, prokhorov2007training}. However, RNN learning is notoriously known to be slow and to be subject to problematic bifurcations \cite{doya1993bifurcations, pearlmutter1995gradient}. To circumvent this problem recent algorithms \cite{pan2012model,waegeman2012feedback} have been based on a reservoir computing architecture and more precisely Echo State Networks (ESN). They are recurrent networks where only the read-out form a random reservoir of neurons is learned \cite{jaeger2001echo}. {In these networks, the slow convergence and the bifurcation issues due to the tuning of the weights are bypassed. However, efficient optimization procedures for the hyper parameters and the overall mathematical understanding of ESNs are still lacking.} Nonetheless, ESN have proved to be very good at handling time dependencies and at predicting time series \cite{jaeger2004harnessing}. Therefore, they provide a solid basis upon which this paper will design a novel control architecture.

Most neural networks for control used so far have been based on two networks: one for the estimation of the state of the environment (which can be called perception), and another for the design of the control (which can be called action). In this paper, I will introduce a somewhat different architecture, since it will only be made of a single recurrent neural network from which two read-outs are drawn and fed back. This is actually a fundamental feature of the approach, which resonates with the field of psychology called ideomotor theory \cite{greenwald1970sensory,shin2010review}. This theory argues that perception and action are tightly linked and even represented in a single "domain". More precisely a fundamental concept is that actions do not aim at changing the world directly, they rather aim at changing the perception of the world. Thus action and perception are deeply entangled and one could say that when the neural network "thinks" or predicts a future for its stimuli, then the corresponding action follows \cite{friston2010action}. {In a way, the 'active inference' perspective presented by Friston and colleagues \cite{friston2010action} could be regarded as a modern theory compatible with ideomotor approach, where the underlying imperative for both action and perception is to minimise prediction errors (or variational free energy).} Alternatively, the conceptual difference from traditional feedback loop algorithms can be seen in the structure of the controller in figure \ref{fig: setup}: at the level of the controller information flows in both directions, what is usually called the output of the controller is also fed-back to the central network.

In this paper, I introduce an ideomotor recurrent neural network (IDRNN) together with a learning procedure in order to control an unknown environment. This paper intends to be a proof of concept that such a neural network can successfully learn to control fairly complicated dynamical systems. In section \ref{sec: model}, I introduce the network and the notations. Section \ref{sec: ideomotor principles} will be devoted to explaining the computational principles underlying ideomotor learning and section \ref{sec: algo} describes the corresponding algorithm. Numerical experiments for various environments and a short comparison with the ESN based method in \cite{waegeman2012feedback} will be presented in section \ref{sec: numerical simus}. Finally, I will discuss the properties of such neural networks, in particular their biological plausibility, in section \ref{sec: discussion}.

\section{Model}\label{sec: model}
The model details the dynamics of a recurrent neural networks interacting with an unknown environment. The way the environment interacts with the agent (through sensors and actuators) is also formally unknown. It is the role of the agent to understand this interaction through statistical observations.
\begin{figure}
 \centering
 \includegraphics[width=0.4\textwidth]{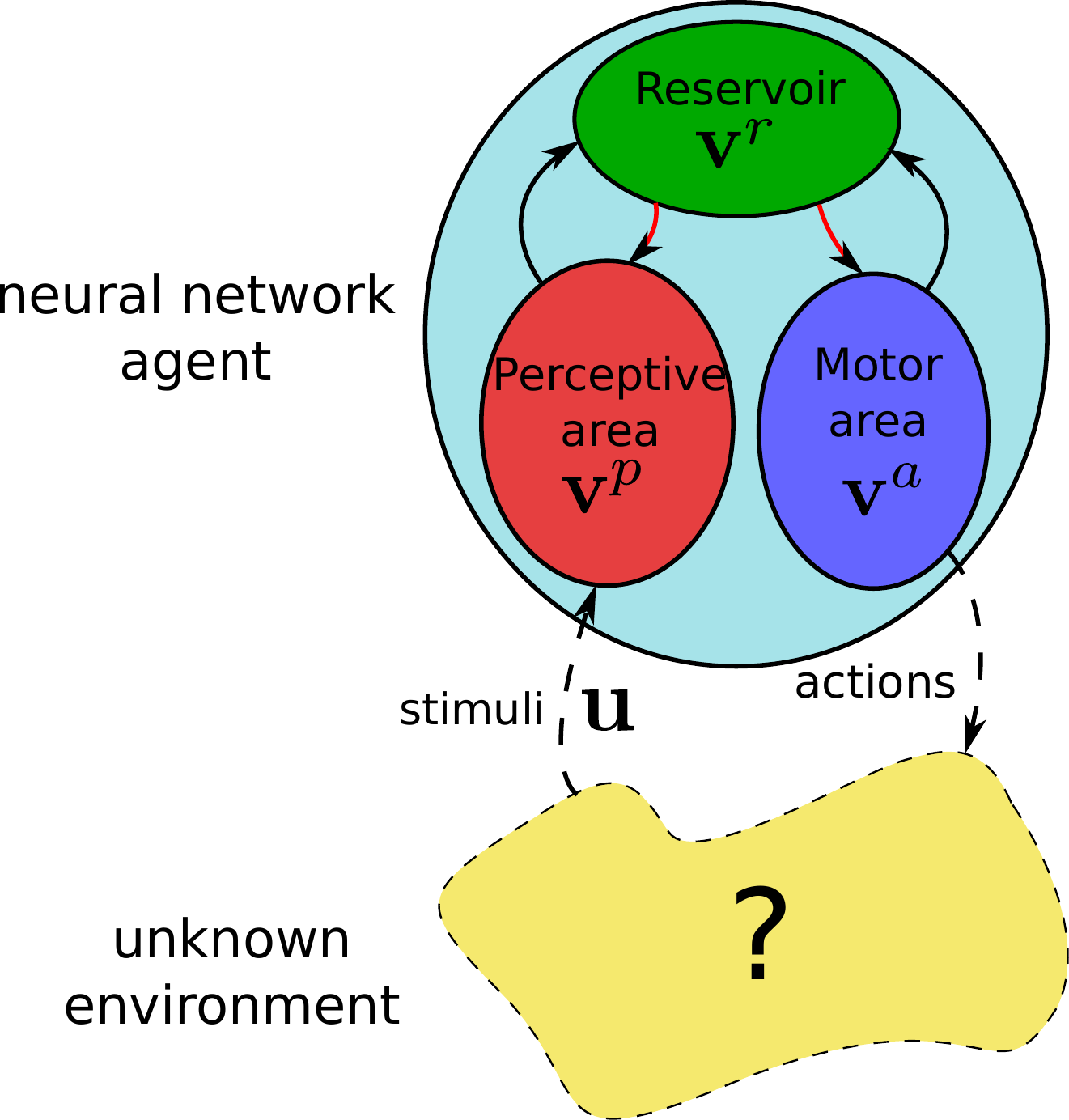}
 \caption{General structure of the proposed controller interacting with its environment. Not all the connections are represented for clarity. Learning only modifies the connections in red (from reservoir to perceptive and motor area).}
 \label{fig: setup}
\end{figure}

\textbf{Notations and conventions}: Several elements of the neural network and the environment are time-dependent multidimensional functions which will be written with lower-case bold letters, e.g. $\v$. The value of these functions at time $t$ will be written as an index to the function's notation, e.g. $\v_t$. Most of these functions will be described by recurrence equations which are assumed to start from $t = 0$ {(for negative time the functions return zero)}. The matrices are written in upper case bold, e.g. $\W$.

We now detail the different parts of the neural network and the environment:
\begin{itemize}
 \item \textbf{Environment}:\\
 The environment possibly has a complicated non-linear and/or stochastic dynamics which we know nothing about. From the controller perspective, only some measures of the environment state are known. They are sometimes called observations but we refer to them as \textit{stimuli} not to confuse them with the states of the reservoir which could also be called observations in the framework of online least mean square problems.
 
The $n_p$ neurons of the perceptive area receive information from the environment through the input vector $\u_t \in \R^{n_p}$ at time $t \in \R_+$ (where $n_p$ is the dimension of the perceptive area).
 \item \textbf{Network activity}:\\
 We assume the controller is made of a neural network, which is decomposed in three parts: a \textit{perceptive area}, a \textit{reservoir} and a \textit{motor area} see, figure \ref{fig: setup}. The variables describing the activity of these subsets of the neural network are respectively $\vp_t \in \R^{n_p}$ called \textit{perception}, $\vr_t \in \R^{\nr}$ {called \textit{reservoir states}} and $\va_t \in \R^{\na}$ called \textit{action}, where $n_p, \nr, \na \in \mathbb{N}^*$ are the number of neurons in the respective area. Note that the letters p, r, a will always stand for perceptive area, reservoir and motor area respectively. We also define the variable gathering the entire network activity: $\v = \begin{pmatrix} \vp & \vr & \va\end{pmatrix}' \in \R^{n_p + \nr + \na}$.
 
 The full connectivity matrix of the network is
 \begin{equation*}
  \W = \begin{pmatrix} 0 & \Wp & 0 \\ \Wrp & \Wrr & \Wra \\ 0 & \Wa & 0 \end{pmatrix}
 \end{equation*}
 {where the second superscript (e.g. b in $\W^{\mbox{ab}}$) is the origin of the connection and the first superscript (e.g. a in $\W^{\mbox{ab}}$) is the destination, consistently with usual matrix notations.} Note that the structure of $\W$ means that there are no connections between and within the perceptive and motor areas.
 
We also define $\Wr = (\Wrp\ \Wrr\ \Wra) \in \R^{\nr \times (\np + \nr + \na)}$  corresponding to all the connections to the reservoir.

A main idea in Echo State Networks \cite{jaeger2001echo} from which this work is inspired, it that the connections within and to the reservoir are randomly drawn. This means that the components of $\Wrr$, $\Wrp$ and $\Wra$ are i.i.d. constants along $\mathcal{N}(0, \sigma^2)$,  $\mathcal{N}(0, \kappa^2)$ and $\mathcal{N}(0, \gamma^2)$. Albeit surprising, this choice leads to relevant prediction results and cheap algorithms.
 
For simplicity, perception and action are considered to be linearly related to the reservoir activity. {The reservoir follows a leaky integrator neural network equation \cite{jaeger2007optimization}, where a time constant $\tau$ controls the speed of the dynamics. Note that the contributions from perceptive and motor areas to the reservoir dynamics are linear (which will be important in the following).} The perceptive area is stimulated by a weighted mean between the stimuli and the current prediction of the network. With the notations introduced before, this reads
\begin{equation}
\left\{
\begin{array}{cl}
 \vr_t  = & (1 - l \tau) \vr_{t-1}  +  \tau \Wr.\begin{pmatrix}s(\vr_{t-1}) \\ \vp_{t-1} \\ \va_{t-1}\end{pmatrix}\\
 \vp_t = & (1-\alpha)\Wp . \vr_t  +   \alpha\u_t\\
 \va_t = & \Wa . \vr_t
\end{array}
\right.
\label{eq: neural network dynamics}
\end{equation}
where $s$ is an element-wise sigmoidal function, $l \in \R_+$ is a decay constant and $\alpha \in [0,1]$ balances the contribution two terms: the stimuli $\u$ and the term $\Wp.\vr_t$ which we call \textit{prediction}. Note that if prediction and stimuli are identical then the perception $\vp$ {becomes the same as the stimuli}.

\item \textbf{Target trajectory}:\\
{The role of the neural network is to control the environment so that it follows a \textit{target} trajectory} which we write $\z_t \in \R^{q}$, where $q\leq \np$. The target corresponds to (at least) one neuron in the perceptive area: if learning is successful, the stimuli to this subset of the perceptive neurons will be driven along the target trajectory. In the present formalism, the neural network can only control its stimuli (as opposed to an unobserved environment state).

\item \textbf{Learning} corresponds to tuning the connections in the network.  We assume that not all the connections are learnable: only the connections in red in figure \ref{fig: setup} are learnable. In other words, and in a reservoir computing spirit, the connections to and within the reservoir $\Wr$ are fixed. We call \textit{perceptive learning} the tuning of the connectivity $\Wp$ and \textit{motor learning} the tuning of the connectivity $\Wa$.
\end{itemize}

\section{Principle of Ideomotor learning}\label{sec: ideomotor principles}
In this paper, ideomotor learning is defined by the combination of two principles: (i) perceptive learning aims at minimizing the distance between internal predictions $\Wp.\vr$ and stimuli $\u$, and, (ii) motor learning aims at minimizing the distance between internal predictions $\Wp.\vr$ and target $\z$.

A control task is said to be successful when stimuli and target are equal. This can be a difficult task to achieve, in particular when the environment is unknown. The idea behind ideomotor learning is that having a good predictor of the stimuli may help designing an intelligent control. A controller which would be able to faithfully reproduce the stimuli, without needing to ``see'' them, would also know how to perturb the environment to reach a desired target.
{This idea is very similar to the good regulator hypothesis (every controller of its environment must possess a model of that environment) that underpins early work in self-organisation and cybernetics \cite{conant1970every}).}  Therefore, one needs to learn a model of the world (principle (i) above) and, at the same time, make sure this model is going in the desired direction (principle (ii) above). More formally, ideomotor learning can be seen as adding a third variable, the prediction, in the distance between stimuli and target and using the triangular inequality to break the minimization of this distance into two sub-tasks.

The fact that motor learning makes no reference to the stimuli $\u$ has important functional consequences. Indeed, the actions exclusively aims at modifying the reservoir dynamics so that the perception matches the target. The impact on the actions on the world and the stimuli $\u$ is simply a byproduct of the method. In fact, the actions only want to control the internal model of the world that perceptive learning tries to build.

{Mathematically, it is possible to formalise the two principles of ideomotor learning as the minimisation of prediction errors by perception and action respectively:}
\begin{equation}
 \begin{array}{lcc}
  \mbox{(i) perceptive learning} & \quad \quad \quad& \underset{\Wp}{\mbox{minimize}} \quad \displaystyle \sum_{s=0}^t \|\u_s - \Wp.\vr_s \|^2\\
  \mbox{(ii) motor learning} & \quad \quad \quad& \underset{\Wa}{\mbox{minimize}} \quad \displaystyle \sum_{s=0}^t \|\z_s - \Wprestr.\vr_s \|^2\\
 \end{array}
 \label{eq: ideomotor principles}
\end{equation}
where $(\u, \vr)$ is a solution of system \eqref{eq: neural network dynamics} defined for given $\Wp$ and $\Wa$. The matrix $\Wprestr \in \R^{q \times \nr}$ is the restriction of the perceptive matrix $\Wp$ to the dimensions which are to be controlled along the trajectory $\z_t$, (i.e. to create $\Wprestr$ some rows of $\Wp$ have been removed). If learning is perfect, i.e. both sums in \eqref{eq: ideomotor principles} are $0$ for all $t$ such that $(\v_t, \u_t)$ is on a limit cycle, then it is clear that the task is reached: $\u_t = \z_t$. We restrict our analysis to the case where such a limit cycle exists (which typically excludes non-controllable, non-observable environments). 

{Note that equation \eqref{eq: ideomotor principles} effectively computes a path integral or time average of squared prediction error or free energy.  As such, equation \eqref{eq: ideomotor principles} expresses a 'principle of least action'; where action is the time average of energy.}

{Designing an online algorithm reaching this limit cycle reveals a fundamental problem: at time step $t$ the network has to figure out new matrices $\Wp$ and $\Wa$ based on the potential impact they would have had in the past. But it cannot really know what this impact would have been since it would need to re-experience the past with these new matrices. This is impossible since the network has to update the matrices exclusively based on the available quantities. Another way to see this is to observe that the values of $\u_t$ and $\vr_t$ depend on $\Wp$ and $\Wa$, and, therefore, learning is not a simple least square problem, where observations $\vr_t$ and target $\u_t$ are independent weight vector. However, simple greedy algorithms can, in certain cases, converge to a perfect solution. A greedy algorithm roughly ignores the dependencies of $\u_t$ and $\vr_t$ on $\Wp$ and $\Wa$, and corresponds to using a traditional least square approach for learning. This is the approach we take in the rest of the paper. {This is formally similar to the mean field approximation that is used in variational formulations of active inference; in other words, minimizing prediction errors under the assumption of conditional independence between the unknown states (and parameters).}}


\section{Greedy RLS algorithm}\label{sec: algo}
In this paper, a greedy Recursive Least Square (RLS) approach to solve the least square problems is considered. It corresponds to dynamically updating the connections based on a RLS algorithm performing the online minimization of the following problem:
\begin{equation}
 \begin{array}{l}
  \underset{\Wp}{\mbox{minimize}} \quad H^p_t = \sum_{s=0}^t \lambdap^{t-s} \|\u_s - \Wp.\vr_s \|^2 + \mup \|\Wp\|^2 \\
  \\
  \underset{\Wa}{\mbox{minimize}} \quad H^a_t = \sum_{s=0}^t \lambdaa^{t-s} \|({\Wprestr}.\Wra)^\dagger.(\z_s - \Wprestr.\vr_s) \|^2 + \mua \|\Wa\|^2 \\
 \end{array}
 \label{eq: RLS pple}
\end{equation}
where $\lambdap, \lambdaa \in [0,1]$ are forgetting factors and $\u$ and $\vr$ are the solution of system \eqref{eq: neural network dynamics} with time dependent connection matrices $\Wp_t$ and $\Wa_t$. If $\lambda$ is close to $0$, then the sum simply involves very recent observations. The additional terms involving the squared norm of $\Wp$ and $\Wa$ correspond to the usual Tikhonov regularization and the numbers $\mup, \mua > 0$ control the amount of regularization. The notation $({\Wprestr}.\Wra)^\dagger$ is the pseudoinverse of ${\Wprestr}.\Wra$.  

The slight modification of the motor criterion in \eqref{eq: RLS pple} due to the inversion of $\Wprestr.\Wra$ simply corresponds to taking a different norm for the minimization. It is necessary to turn motor learning into a classical weighted least square problem. Indeed, one can unravel the dynamics of the reservoir for one time step to let the connection explicitly appear in the criterion. This corresponds to injecting the first row of \eqref{eq: neural network dynamics} into $\z_t - \Wprestr.\vr_t$ which leads to
\begin{equation}
\z_t - \Wprestr.\vr_t = \tau(\y_t -  \Wprestr.\Wra.\Wa.\vr_{t-1})
\end{equation}
where $\y_t = \frac{\z_t - \Wprestr.(1 - l \tau) \vr_{t-1}}{\tau}  -  \Wprestr.\Wrr.s(\vr_{t-1}) -  \Wprestr.\Wrp.\vp_{t-1}$. As a consequence, $\z_t - \Wprestr.\vr_t$ appears as a linear function of $\Wa$ which will make it possible to use classical algorithms for minimization. Note that the particular dynamics of the reservoir in \eqref{eq: neural network dynamics} was chosen such that such a linear problem would appear. It also explains why we cannot unravel the dynamics for more time steps: the reformulation into a least square problem would be impossible. From this formulation, it becomes clear that pre-multiplying the factor $\z_t - \Wprestr.\vr_t$ by $({\Wprestr}.\Wra)^\dagger$ for the motor learning with the RLS algorithm in equation \eqref{eq: RLS pple} is useful. Indeed, motor learning  becomes the minimization of
\begin{equation}
H^a_t = \sum_{s=0}^t \lambdaa^{t-s}\|({\Wprestr}.\Wra)^\dagger.\y_s - \Wa.\vr_{s-1}\|^2 + \mua \|\Wa\|^2
\label{eq: modified ideomotor principles}
\end{equation}
which takes the form of a classic weighted least square problem which can directly be solved by a RLS algorithm. Note that matrix $\Wprestr.\Wra$ has the size of the number of controlled dimensions (spatial dimension of the target) times the number of dimension for action, which is likely to be small enough to enable a computationally cheap inversion at each time step.

The greedy RLS algorithm consists in implementing \eqref{eq: RLS pple} with a RLS algorithm which quickly forgets the past. Indeed, to circumvent the dependency of $\u$ and $\Wp.\vr$ on $\Wa$ the choice of the forgetting factors $\lambdap, \lambdaa$ is crucial. When the network starts from an uninformed state (e.g. the null state), it is important for them to have small values, typically $0.99$. A rule of thumb \cite{haykin2005adaptive} to tune them is that the memory of the algorithm roughly corresponds to $\frac{1}{1 - \lambda}$ time steps.

\paragraph{The regularized RLS algo} is recalled in this paragraph. It is an algorithm which recursively solves the following generic problem
\begin{equation}
\underset{\W_t}{\mbox{minimize}} \quad H_t = \sum_{s=0}^t \lambda^{t-s} \|\x_s - \W_t.\y_s \|^2 + \mu \|\W\|^2
\end{equation}
It can be solved by iterating the following regularized RLS step \cite{haykin2005adaptive, gunnarsson1996combining} as new targets $\x$ and new observation $\y$ arrive. This algorithm has already been successfully used for Echo State Networks \cite{jaeger2004harnessing, sussillo2009generating, laje2013robust}. {In this paper, we use a version of the algorithm with non-fading regularization \cite{gunnarsson1996combining}, which provides good stability properties even when the forgetting factor $\lambda$ is small.} The algorithm corresponding to one step of the considered RLS algorithm is given in algorithm \ref{alg: RLS step}.
\begin{algorithm}[htbp]
\caption{Regularized RLS step}\label{alg: RLS step}
\begin{algorithmic}[1]
\Procedure{rls$\_$step}{$\P$, $\W$, $\x$, $\y$, $\mu$, $\lambda$}
\State $\P \gets \P - \mu \P^2$
\State $\P \gets \frac{1}{\lambda}\left(\P -  \frac{\P.\y.\y'.\P}{\lambda + \y'.\P.\y}\right)$
\State $\W \gets \W + (\x - \W.\y).(\P.\y)'$
\State return $\P$, $\W$
\EndProcedure
\end{algorithmic}
\end{algorithm}
In this algorithm, $\P \in \R^{n \times n}$ is the inverse correlation matrix of the observations, which correspond to the reservoir activity in this paper.

\paragraph{Pseudo-code}
Therefore, the algorithm summarizing the update of the greedy RLS neural network is described in algorithm \ref{alg: full RLS agent}.
\begin{algorithm}[htbp]
\caption{IDRNN}\label{alg: full RLS agent}
\begin{algorithmic}[1]
\State $\#$ Initialization:
\State $\vp, \vr, \va, \Wa \gets 0$
\State $\Wp, \Wr_{ij} \gets \mathcal{N}(0,*)$
\State $\P^p, \P^a \gets \eta I_d$
\State $\#$ Main loop:
\While{$\u \gets$ new stimuli \textbf{and} $\z \gets$ new target}
\State \# Motor learning:
\State $\y \gets ({\Wprestr}.\Wra)^\dagger.\left(\z - (1 - l \tau) \Wprestr.\vr - \tau \Wprestr.\Wrr.s(\vr) - \tau \Wprestr.\Wrp.\vp\right)$
\State $\P^a, \Wa \gets  \mbox{RLS}\_\mbox{STEP}(\P^a, \Wa, \y, \vr,\mua, \lambdaa)$
\State \# Reservoir update:
\State $\vr  \gets (1 - l \tau) \vr  +  \tau \Wr.\begin{pmatrix}\vp & s(\vr) & \va \end{pmatrix}'$
\State \# Perceptive learning:
\State $\P^p, \Wp \gets  \mbox{RLS}\_\mbox{STEP}(\P^p, \Wp, \u, \vr,\mup, \lambdap)$
\State \# Perception and action update:
\State $\vp \gets (1-\alpha)\Wp . \vr  +   \alpha\u$
\State $\va \gets \Wa . \vr$ is the action which controls the environment.
\EndWhile
\end{algorithmic}
\end{algorithm}

\section{Numerical experiments}\label{sec: numerical simus}
In this section, I show on different examples that the proposed architecture can effectively control an unknown environment along a desired trajectory. It is important to realize that the following numerical simulations correspond to a neural network which learns to control the environment: at time $t=0$ the neural network has no clue about what system it is dealing with. Thus the results are not to be compared to a classical control setup where the controller was previously tuned to the specific environment.

\subsection{Learning to control random neural networks}
As toy models of the environment, I first choose randomly connected neural networks from which a linear read-out is to be controlled along a sinus function. It has been argued that this class of systems is very rich since it can approximate any dynamical system \cite{sontag1997recurrent}. I do not claim that the IDRNN can control any such random neural network since they can become very complicated, but it performs well on reasonably complicated instances of such networks. I will consider neural networks made of $5$ neurons from which a one-dimensional linear readout is computed. The equation governing these driven random neural networks is:
\begin{equation*}
\left\{
\begin{array}{rl}
\x_{t+1} & = (1 - \bar{l} \bar{\tau}) \x_t + \tau \bar{\W}.\tanh(\x_t) + \b.\va_t\\
\u_t & = \c'.\x_t
\end{array}
\right.
\end{equation*}
where matrix $\bar{\W} \in \R^{5 \times 5}$ and the vectors $\b, \c \in \R^5$ have components which are drawn i.i.d. $\mathcal{N}(0, \bar{\sigma}^2)$,  $\mathcal{N}(0, \bar{\kappa}^2)$ and $\mathcal{N}(0, \bar{\gamma}^2)$ respectively.

\begin{figure}[htbp]
	\centering
		\subfigure[Freely run environment.]{\includegraphics[width=.45\textwidth]{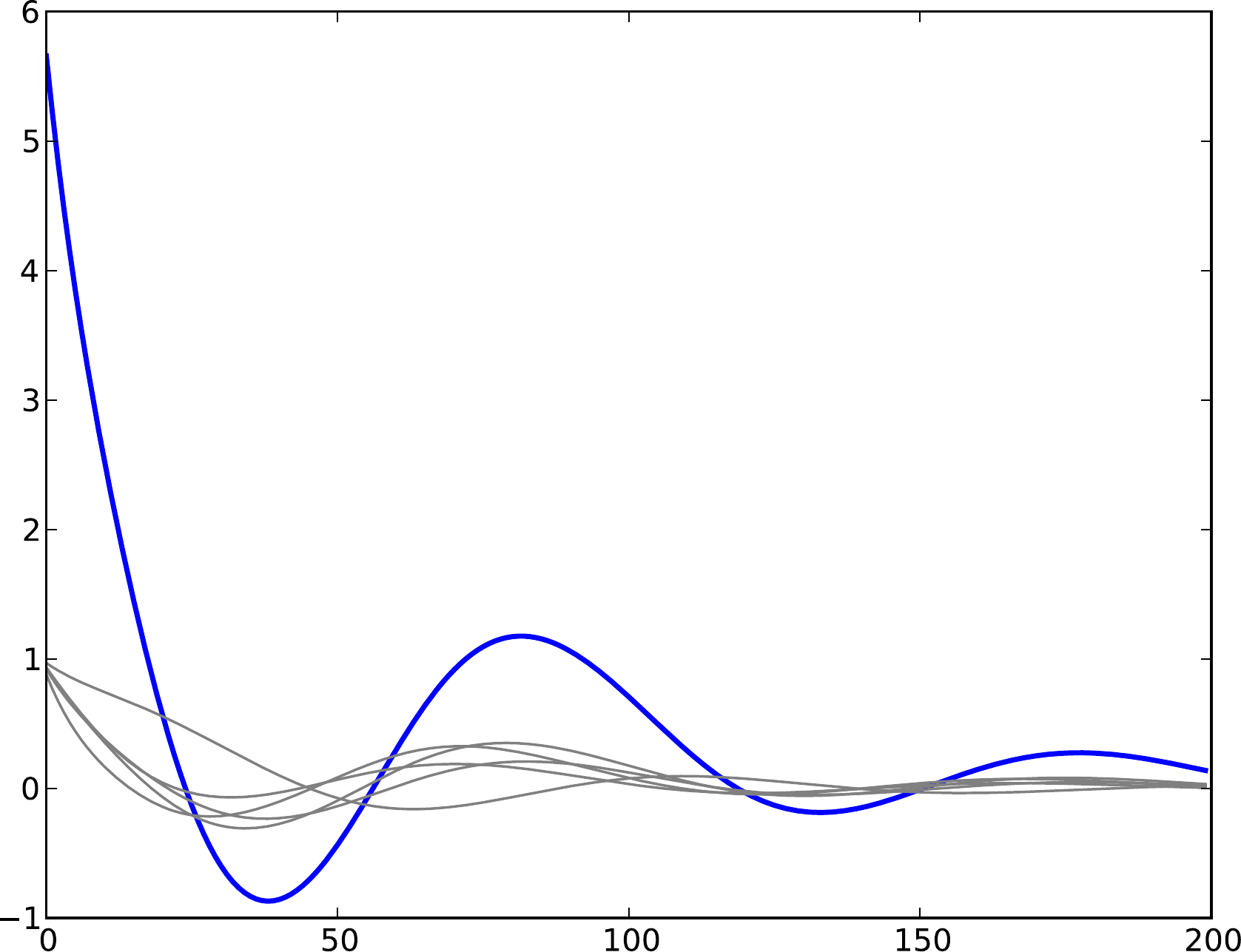} \label{fig: RNN conv free}}\qquad
		\subfigure[Controled environment.]{\includegraphics[width=.45\textwidth]{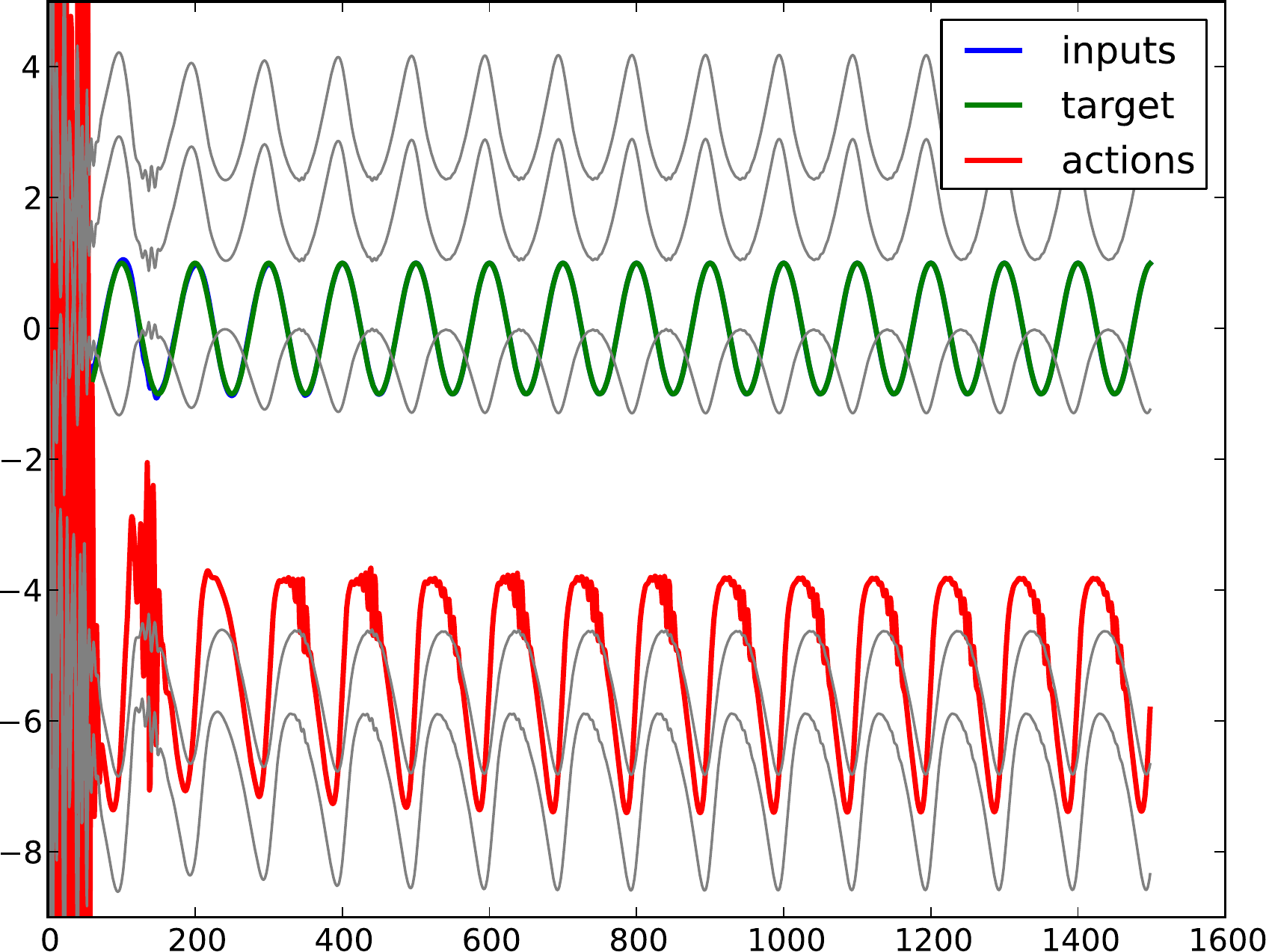} \label{fig: RNN conv contr}}
	\caption{Learning to control a converging random neural network. In grey are represented the 5 components of the environment states $\x_t$. In blue is the read-out from the environment $\u_t$ which corresponds to the stimuli or inputs to the IDRRN. In red are the actions $\va_t$ of the IDRNN on the environment. Environment parameters: $\bar \sigma =\bar \kappa = \bar \gamma = 1$, $\bar l = 1$, $\bar \tau = 0.1$.}
	\label{fig: RNN conv}
\end{figure}

\begin{figure}[htbp]
	\centering
		\subfigure[Freely run environment.]{\includegraphics[width=.45\textwidth]{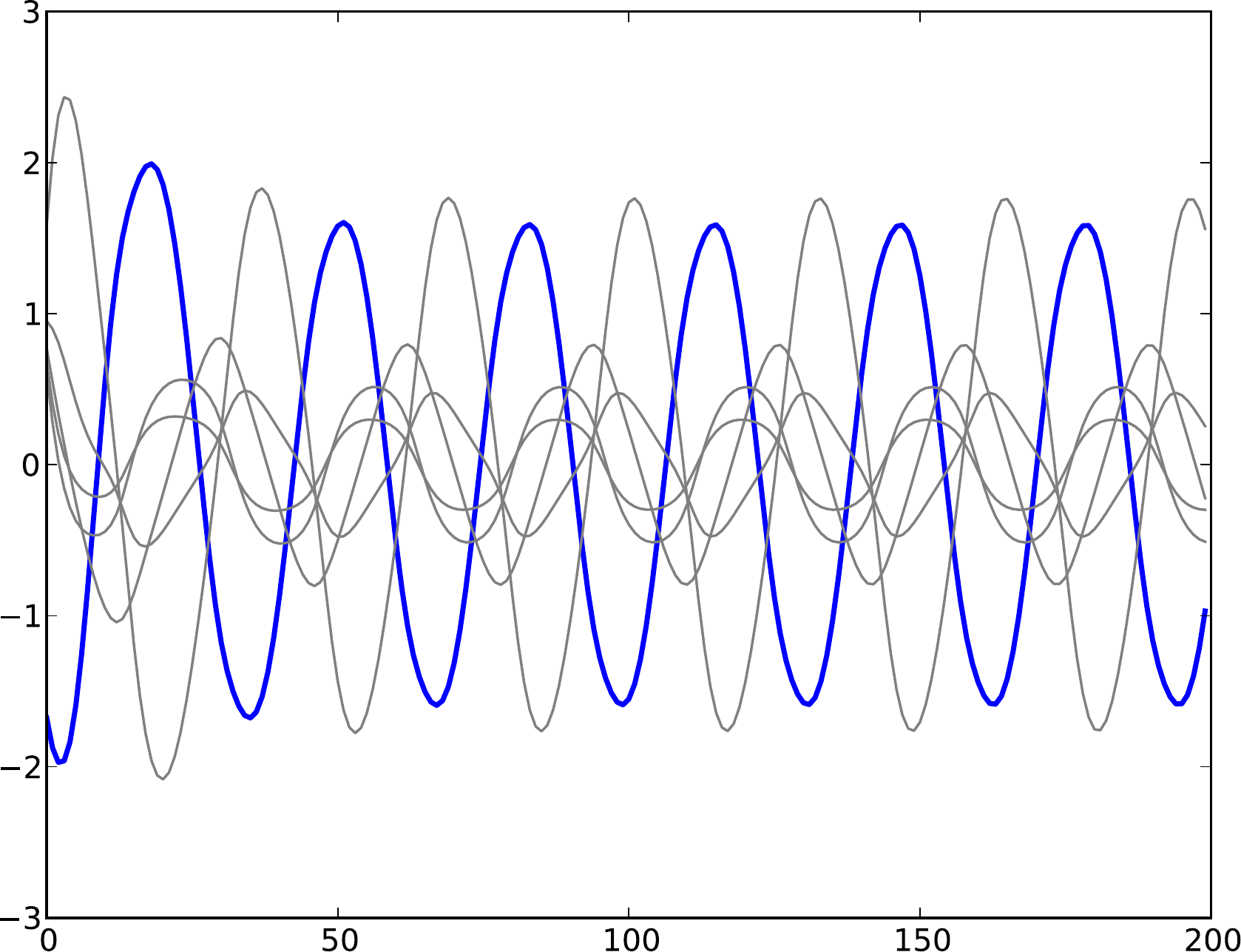} \label{fig: RNN osc free}}\qquad
		\subfigure[Controlled environment.]{\includegraphics[width=.45\textwidth]{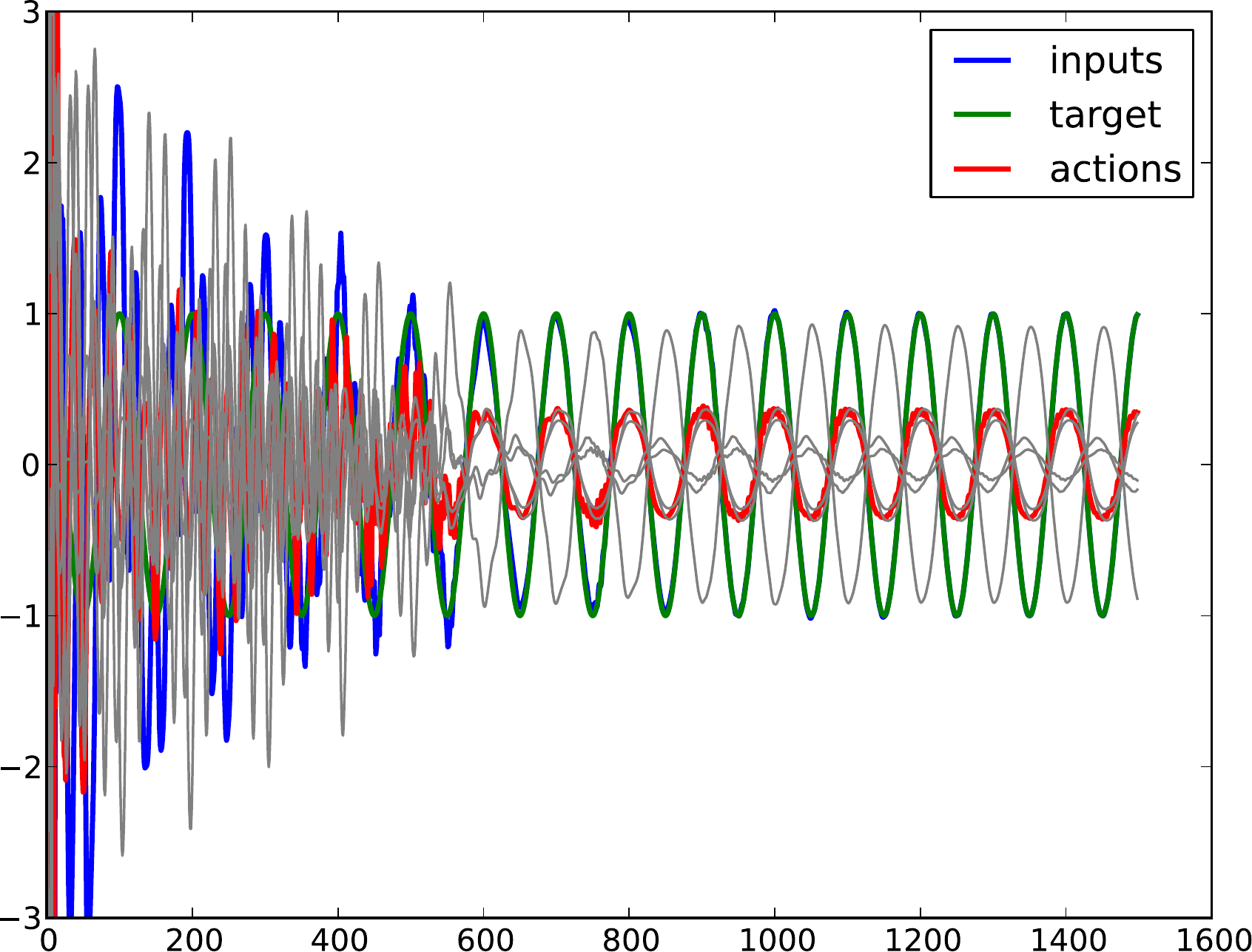} \label{fig: RNN osc contr}}
	\caption{Learning to control an oscillatory random neural network. See figure \ref{fig: RNN conv} for the color code. Environment parameters: $\bar \sigma =\bar \kappa = \bar \gamma = 1$, $\bar l = 1$, $\bar \tau = 0.15$.}
	\label{fig: RNN osc}
\end{figure}

\begin{figure}[htbp]
	\centering
		\subfigure[Freely run environment.]{\includegraphics[width=.45\textwidth]{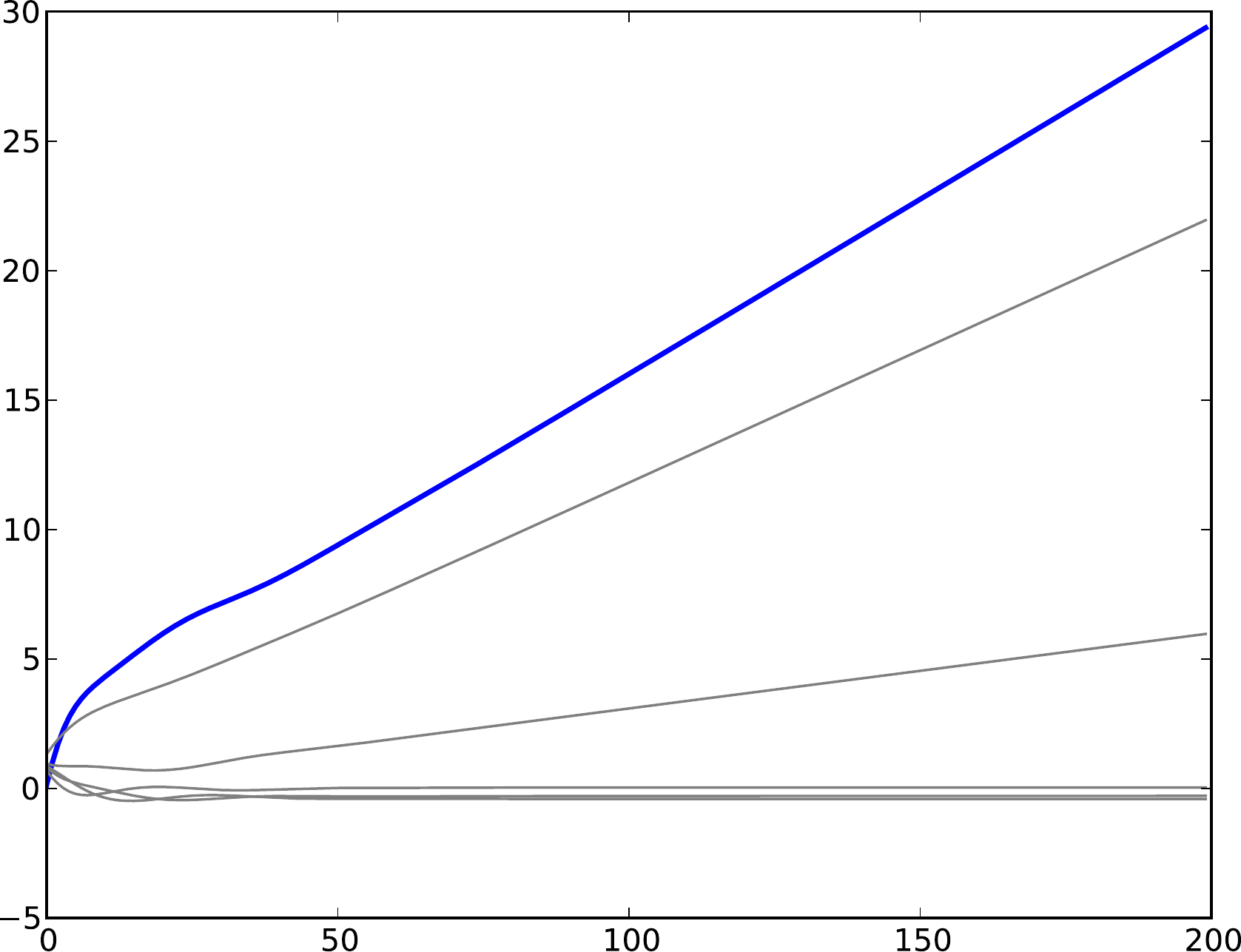} \label{fig: RNN div free}}\qquad
		\subfigure[Controled environment.]{\includegraphics[width=.45\textwidth]{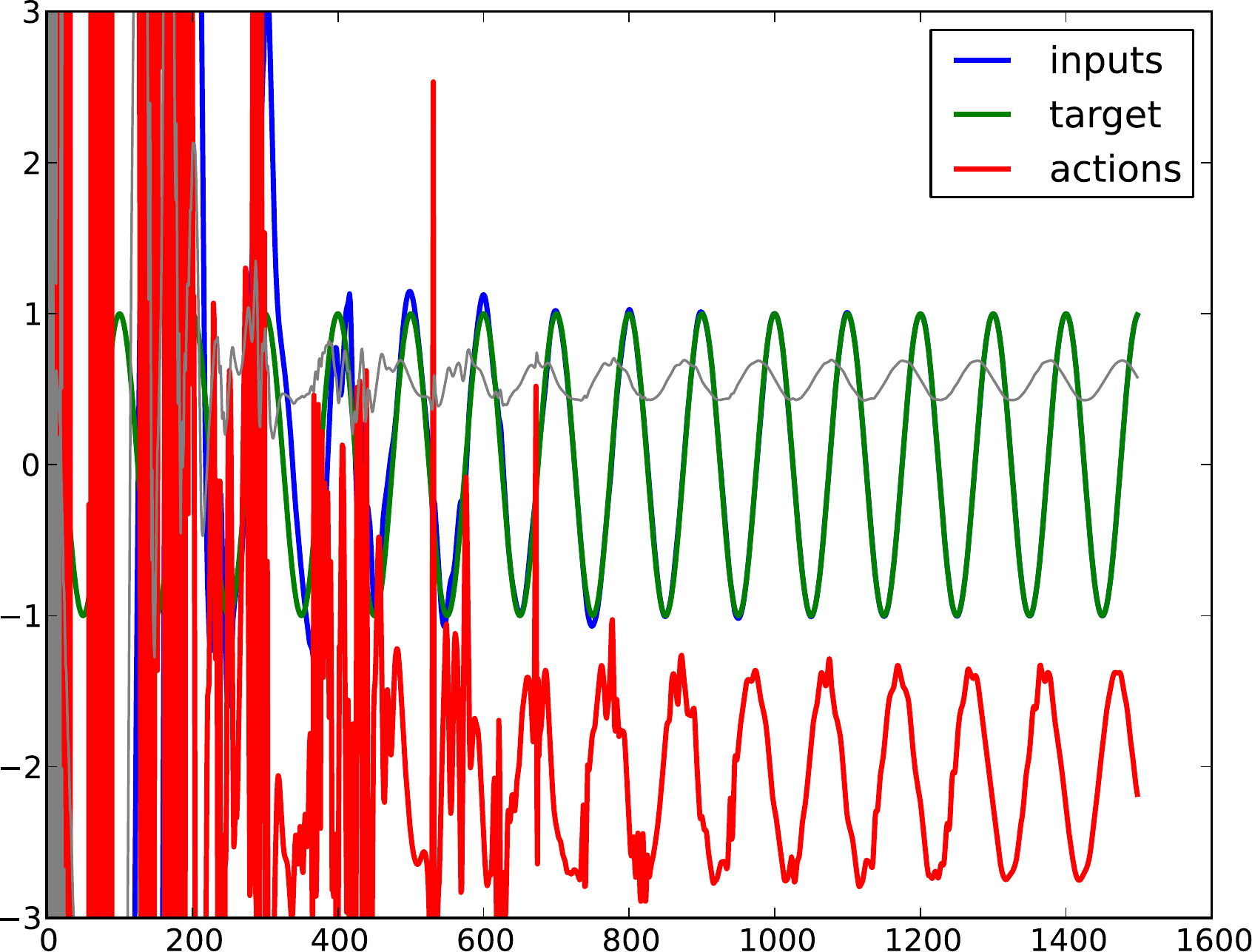} \label{fig: RNN div contr}}
	\caption{Learning to control a diverging random neural network. See figure \ref{fig: RNN conv} for the color code. Environment parameters: $\bar \sigma =\bar \kappa = \bar \gamma = 1$, $\bar l = 0$, $\bar \tau = 0.1$.}
	\label{fig: RNN div}
\end{figure}

With different choices of parameters, one can get very different resulting dynamics as shown in figures \ref{fig: RNN conv free}, \ref{fig: RNN osc free}, \ref{fig: RNN div free}. Even with identical parameters the different realizations of the connections can lead to qualitatively different dynamics. I chose three cases corresponding to  converging (figure \ref{fig: RNN conv}), oscillatory (figure \ref{fig: RNN osc}) and diverging (figure \ref{fig: RNN div}) situations.

For all the simulations, I chose the same parameters for the IDRNN: $\na = \np = 1$, $\nr = 100$, $\sigma = 1.5$, $\kappa = \gamma = 0.1$, $l = 1$, $\tau = 0.1$, $\alpha = 0.5$, $\eta = 1$, $s(x) = \tanh(x + 0.01)$, $\lambdap = 0.99$, $\lambdaa = 0.9$, $\mup = 10^{-6}$, $\mua = 10^{-3}$. {The motivation for the choice of $\sigma>1$ and $s(0) \neq 0$ is to have a reservoir with spontaneous activity and no equilibrium point at $\vr =0$ when no stimulation comes in.} The parameters of the environment are reported in the figures.

In each situation, the very same neural network has managed to control qualitatively different environments it knew nothing about along the same trajectory $z_t = \sin(2 \pi t / 100)$. The $L_1$ error between stimuli and target over the last $100$ time steps (corresponding to a period of the target) is respectively $0.0006$, $0.0046$ and $0.0017$ for converging, oscillatory and diverging situations, meaning that the control is successful.

In each case, the first few hundred time steps display an extremely variable behavior. The network has difficulties choosing relevant actions since its perception is not fully developed yet. Although the variations look completely unstructured the network still manages to calm down and converges to the desired trajectory. Unfortunately, this is not always the case: the search period can bring the environment to an undesired location in the state space leading to a failure of the control. For instance, controlling a quickly diverging environment (not shown) can fail when the environment diverges faster than the network can learn. The two important parameters to control the behavior of the network in the beginning are the initial values of $\Wp$ and $\P_p, \P_a$. If the value of $\Wp_0$ is to small then the pseudo-inversion of $\Wprestr.\Wra$ in the algorithm can lead to extremely large control values in the very beginning of the simulation. Similarly, choosing large initial values for $\P_p$ or $\P_a$ induces a large variability at the beginning of the simulation \cite{haykin2005adaptive} which can lead to poor results.

Observe in figure \ref{fig: RNN div contr} that the hidden variables in the environment $\x$ are not displayed. Indeed, although the network is controlling properly the environment read-out $\u$ it does not necessarily prevent these hidden variables from diverging. This leads to a numerical overflow in the simulation after a certain time and an increased variability due to numerical round-offs. To prevent this one would need to ask the IDRNN to control also the states $\x$.

\subsection{Learning to control delayed systems}
Actually, the IDRNN in its current form fails at controlling delayed systems or even some instantaneous systems which for which the impact of an action may take some time steps to reveal itself. This is not a surprise since the network only tries to control what it happening from one time step to the other.

A simple extension of the algorithm \ref{alg: full RLS agent} makes the IDRNN able of controlling such delayed systems. The idea is to learn to predict the future of the stimuli at $t + \delta$ time steps with $\delta > 1$. This corresponds to  adding a third readout matrix $\Wf$ to the network in figure \ref{fig: setup}. The predicted value $\Wf.\vr_t$ should be a approximation of $\u_{t + \delta}$ and is not fed back to the network. In practice, learning such a future prediction matrix can be done by applying an RLS algorithm to approximate $\u_t$ based on the observation $\vr_{t-\delta}$. This implies to store the history of the reservoir during $\delta$ time steps. Once this prediction is set-up it suffices to replace $\Wprestr$ by $\Wf$ in algorithm \ref{alg: full RLS agent}. This leads to a delayed version of the IDRNN detailed in algorithm \ref{alg: delayed RLS agent} which does have a time window similar to \cite{waegeman2012feedback}.

\begin{algorithm}[htbp]
\caption{Delayed IDRNN}\label{alg: delayed RLS agent}
\begin{algorithmic}[1]
\State $\#$ Initialization:
\State $\vp, \vr, \va, \Wp, \Wa \gets 0$
\State $\Wf, \Wr_{ij} \gets \mathcal{N}(0,*)$
\State $\P^p, \P^a, \P_f \gets \eta I_d$
\State $\#$ Main loop:
\While{$\u \gets$ new stimuli \textbf{and} $\z \gets$ new target}
\State \# Motor learning:
\State $\y \gets ({\Wf}.\Wra)^\dagger.\left(\z - (1 - l \tau) \Wf.\vr - \tau \Wf.\Wrr.s(\vr) - \tau \Wf.\Wrp.\vp\right)$
\State $\P^a, \Wa \gets  \mbox{RLS}\_\mbox{STEP}(\P^a, \Wa, \y, \vr,\mua, \lambdaa)$
\State \# Reservoir update:
\State $\vr  \gets (1 - l \tau) \vr  +  \tau \Wr.\begin{pmatrix}\vp & s(\vr) & \va \end{pmatrix}'$
\State $\vr_{\mbox{list}} \gets [\vr, \vr_{\mbox{list}} ]$
\State \# Perceptive learning:
\State $\P^p, \Wp \gets  \mbox{RLS}\_\mbox{STEP}(\P^p, \Wp, \u, \vr,\mup, \lambdap)$
\State $\P^f, \Wf \gets  \mbox{RLS}\_\mbox{STEP}(\P^f, \Wf, \u, \vr_{\mbox{list}}(\delta),\muf, \lambdaf)$
\State \# Perception, future and action update:
\State $\vp \gets (1-\alpha)\Wp . \vr  +   \alpha\u$
\State $\vf \gets \Wf.\vf$ is {a proxy of the future inputs}.
\State $\va \gets \Wa . \vr$ is the action which controls the environment.
\EndWhile
\end{algorithmic}
\end{algorithm}

With this new algorithm, it is possible to control systems for which the instantaneous IDRNN failed. Here I consider two examples of such systems: first a linear oscillatory system with a single control variable and, second, a random neural network with delayed action.

\begin{figure}[htbp]
	\centering
		\subfigure[Freely run environment.]{\includegraphics[width=.45\textwidth]{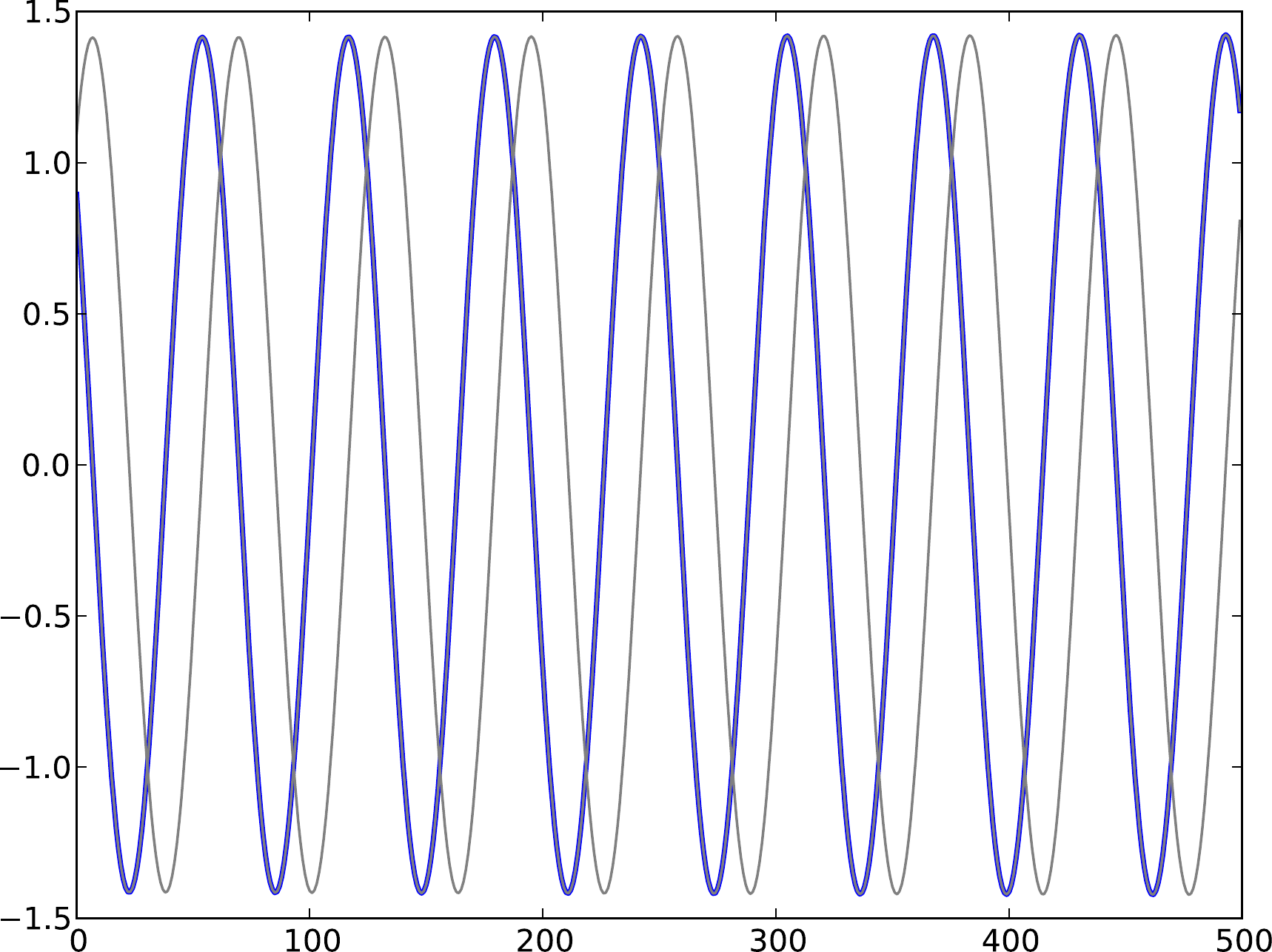} \label{fig: linear osc free}}\qquad
		\subfigure[Controled environment.]{\includegraphics[width=.45\textwidth]{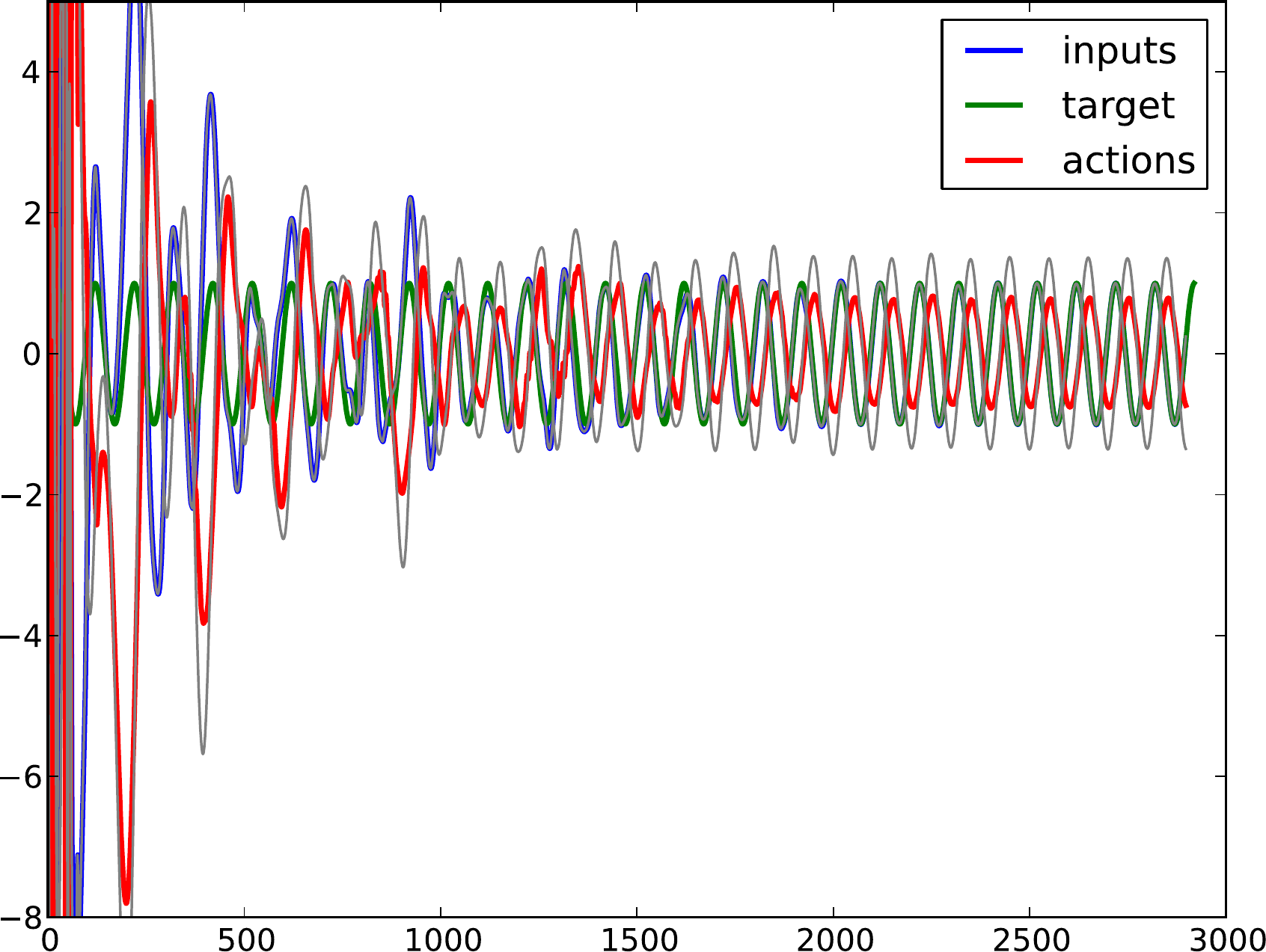} \label{fig: linear osc contr}}
	\caption{Learning to control a linear oscillatory system with a single positive control variable. See figure \ref{fig: RNN conv} for the color code.}
	\label{fig: linear osc}
\end{figure}

In a first time, we consider the following environment to control
\begin{equation*}
\begin{pmatrix}\u_{t+1} \\ \x_{t + 1} \end{pmatrix} =
\begin{pmatrix}\u_{t} \\ \x_{t} \end{pmatrix} + 
0.1\begin{pmatrix}-0.05 & -1 \\ 1 & -0.05 \end{pmatrix}.\begin{pmatrix}\u_{t} \\ \x_{t} \end{pmatrix} + 0.1
\va_t \begin{pmatrix}1 \\ 0.5 \end{pmatrix}
\end{equation*}
This corresponds to a linear oscillatory system as shown in figure \ref{fig: linear osc free} somewhat similar to the first system studied in \cite{waegeman2012feedback}. The negative terms on the diagonal compensate for the tendency of the Euler algorithm to diverge when simulating linear systems and can be disregarded for the interpretation. Actually, the difficulty to control the system comes from the control vector $(1\ 0.5)$ which is positive for both components. To understand intuitively the problem it is useful to see that one dimension is excitatory while the second is inhibitory. When the control variable only makes it possible to positively stimulate both dimensions, it is not clear what will be the situation in a few time steps: although the excitatory dimension is directly stimulated by the inputs, the inhibitory dimension will suppress this excitation (even more since it is stimulated by the excitatory dimension). In the end, one needs to see a bit more in the future to understand the impact of the action. This is precisely why adding the time window $\delta$ to the prediction makes the task much more simple. Indeed, the IDRNN can effectively control the system as shown in figure \ref{fig: linear osc contr}. The parameters used are the same as previously with the addition of $\delta = 20$, $\lambdaf = 0.995$ and $\eta =10^{-4}$.

The second environment is governed by a random neural network as previously but with a delayed command:
\begin{equation*}
\left\{
\begin{array}{rl}
\x_{t+1} & = (1 - \bar{l} \bar{\tau}) \x_t + \tau \bar{\W}.\tanh(\x_t) + \b.\va_{t - 50}\\
\u_t & = \c'.\x_t
\end{array}
\right.
\end{equation*}
The parameters of the IDRNN are identical to before except that $\delta = 55$, $\lambdaa$ and $\eta = 10^4$ which explains the great variability at the beginning of the simulation.

\begin{figure}[htbp]
	\centering
		\subfigure[Freely run environment.]{\includegraphics[width=.45\textwidth]{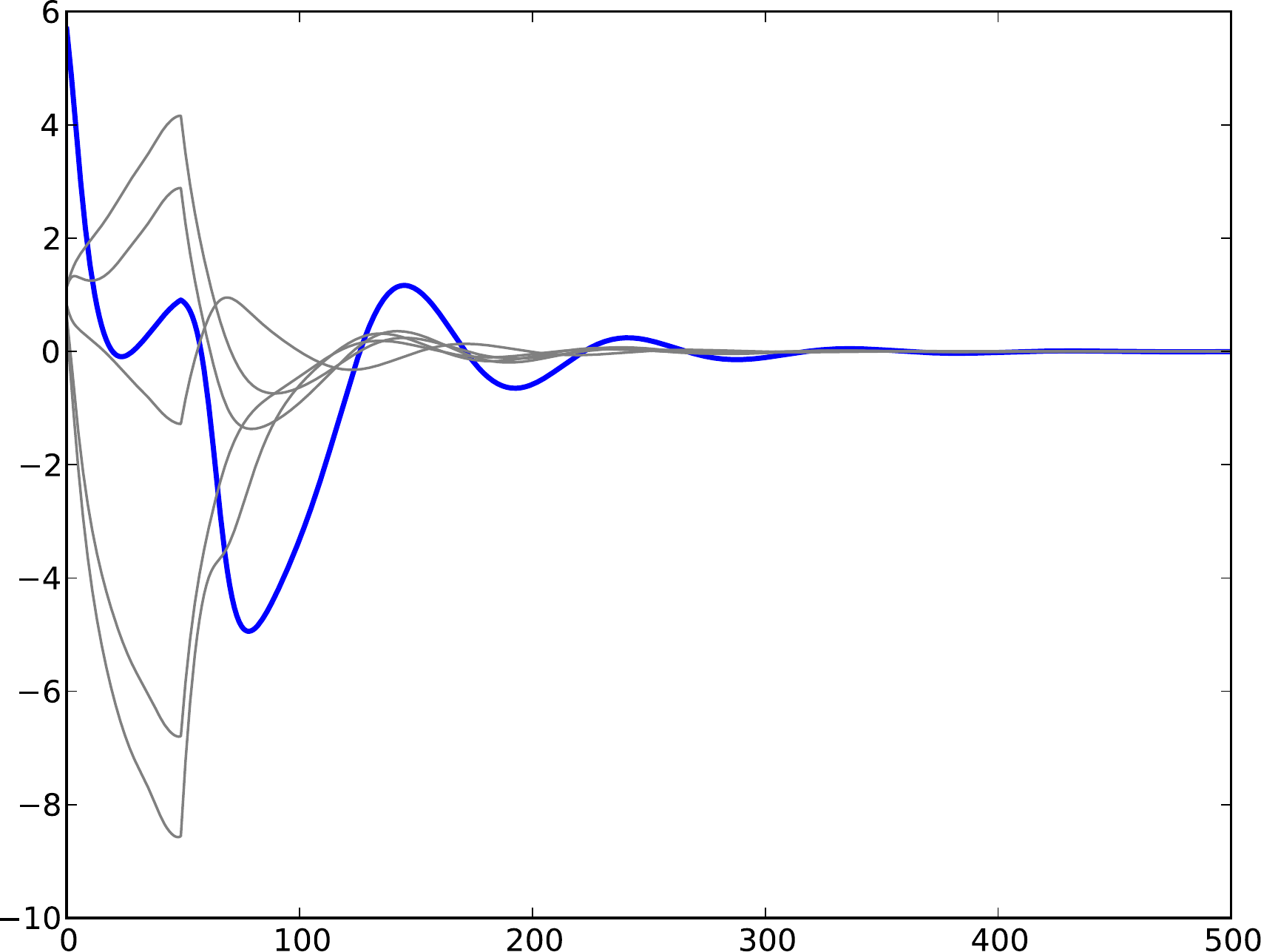} \label{fig: RNN del free}}\qquad
		\subfigure[Controled environment.]{\includegraphics[width=.45\textwidth]{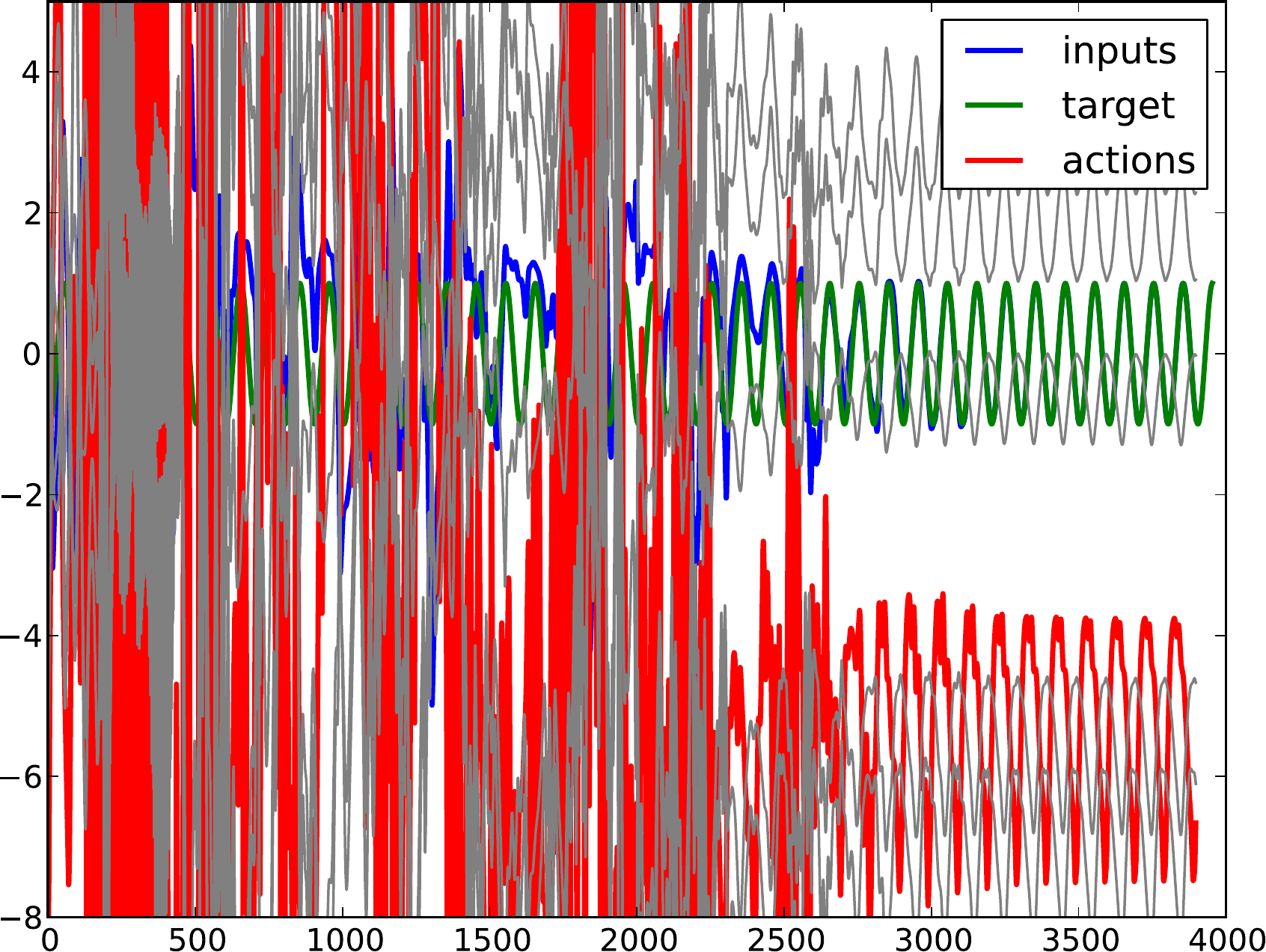} \label{fig: RNN del contr}}
	\caption{Learning to control of a random neural network with a delayed control variable. See figure \ref{fig: RNN conv} for the color code. Environment parameters: $\bar \sigma =\bar \kappa = \bar \gamma = 1$, $\bar l = 1$, $\bar \tau = 0.1$.}
	\label{fig: RNN del}
\end{figure}

The control performed by the IDRNN is displayed in figure \ref{fig: RNN del}. Although the control takes more time ($4000$ time steps instead of $1500$), it finally manages to bring the system along the desired trajectory.

\subsection{Short comparison with the double reservoir architecture}
Although an extensive comparison between the IDRNN and the double reservoir architecture (DRA) \cite{waegeman2012feedback} is beyond the scope of this paper, I show here that they perform similarly on the task of controlling a heating tank. 

The system to control, a heating tank, is the same as the second example in \cite{waegeman2012feedback} and the interested reader should refer to this paper for the implementation details (I took the same parameters). Briefly, the system is a non-linear state-delayed system where the delay depends on the action. It corresponds to controlling a heating tank whose input is a stream of cold water. The water is heated in the tank and is then channeled through a tube to a destination where the output temperature is measured. When the water throughput (corresponding to the action $\va$) is increased the output temperature decreases, but it takes less time to get out of the tube. It is shown in \cite{waegeman2012feedback}, that the DRA outperforms another algorithm, called NEPSAC \cite{galvez2009nonlinear}, on this task.

\begin{figure}[htbp]
	\centering
		\subfigure[Target trajectory $\z$.]{\includegraphics[width=.45\textwidth]{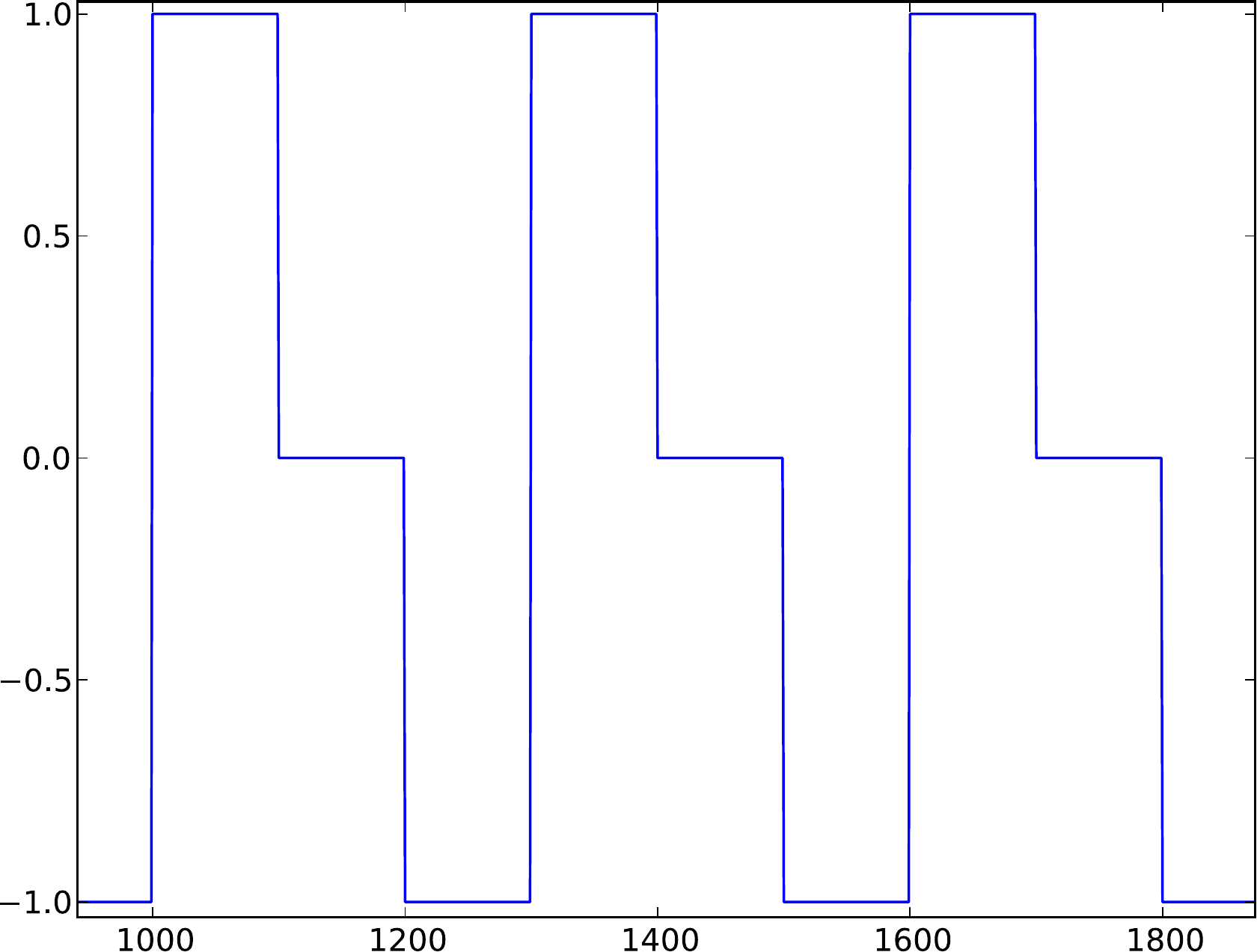} \label{fig: target}}\qquad
		\subfigure[Initialization of IDRNN.]{\includegraphics[width=.45\textwidth]{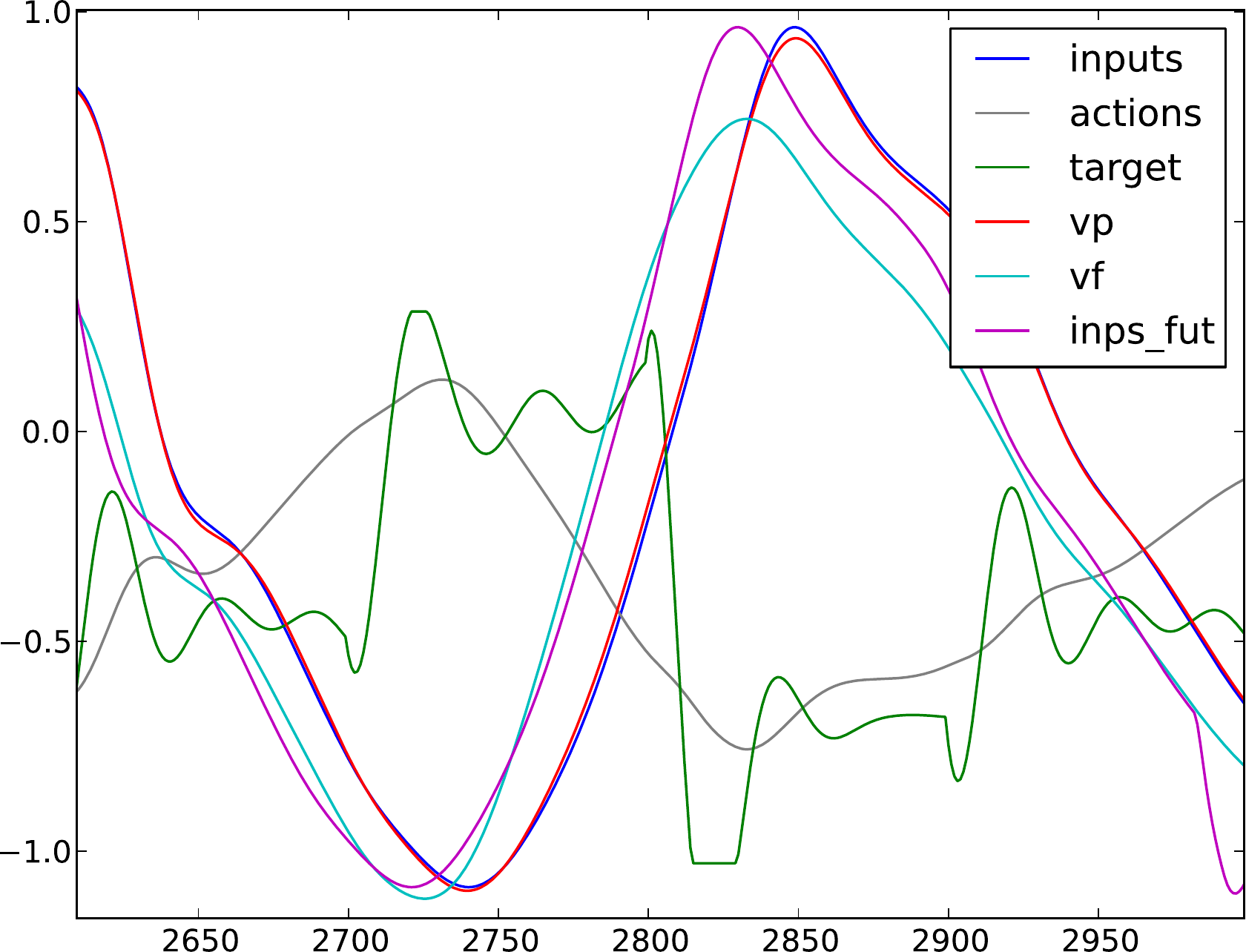} \label{fig: IDRNN init}}
	\caption{(left) target trajectory used for the heating tank experiment. (right) Result of the initialization of the IDRNN. The actual implementation of algorithm \ref{alg: delayed RLS agent} started right after. In blue are the inputs $\u_t$ and in pink are the future inputs $\u_{t + \delta}$. In grey are the actions $\va_t$. In green is the target for the initialization phase (different from $\z_t$). Note that it is a target for the action and not the perception. In red is the perception $\vp$ and in cyan in the prediction of the future $\vf$.}
	\label{fig: heating tank setup}
\end{figure}

In this paper, there are two differences from the heating tank used in \cite{waegeman2012feedback}: first, the stimuli to the network are scaled to meet the networks dynamics. I take as input $\frac{T - 30}{5}$ where $T$ is the output of the heating tank simulation. Second, the target trajectory is different, see figure \ref{fig: target}. It is much faster so that the results of the control algorithm do not look as good as in the original paper presenting the DRA. I have not tried to tune the parameters to get the best results, since I only intended to show that both algorithm would perform similarly on this task with naive parameter tuning.

Actually, the IDRNN failed to control the system when it is randomly initialized. Interestingly the DRA could control the system although I did not use any babbling initialization as discussed in \cite{waegeman2012feedback}. To get good results with the IDRNN I had to properly initialize the network. {Indeed, a useful property of the IDRNN is its ability to be initialized along any prexisting control scheme. The initialization procedure consists in recording both perceptions and actions (more commonly named inputs and outputs) of another control scheme (here I took the DRA) during a long time (here I took $3000$ time steps). Then a simple ESN procedure can be used to reproduce the recorded trajectories by learning the appropriate feedback connections $\Wp$ and $\Wa$. After learning, when the very same environment is presented again to the neural network, it will reproduce both predictions and actions of the learned control scheme. However, if the learning time was to small or the environment variable enough, then  there may be some discrepancies between recorded and reproduced trajectories, which is the case in this application:} at the end of initialization the system had a behavior which was very different from the DRA, see figure \ref{fig: IDRNN init}. Nonetheless it was sufficient for the coupled system to be in an attractor {basin} and the control did work.

\begin{figure}[hbtp]
\centering
\includegraphics[width=.6\textwidth]{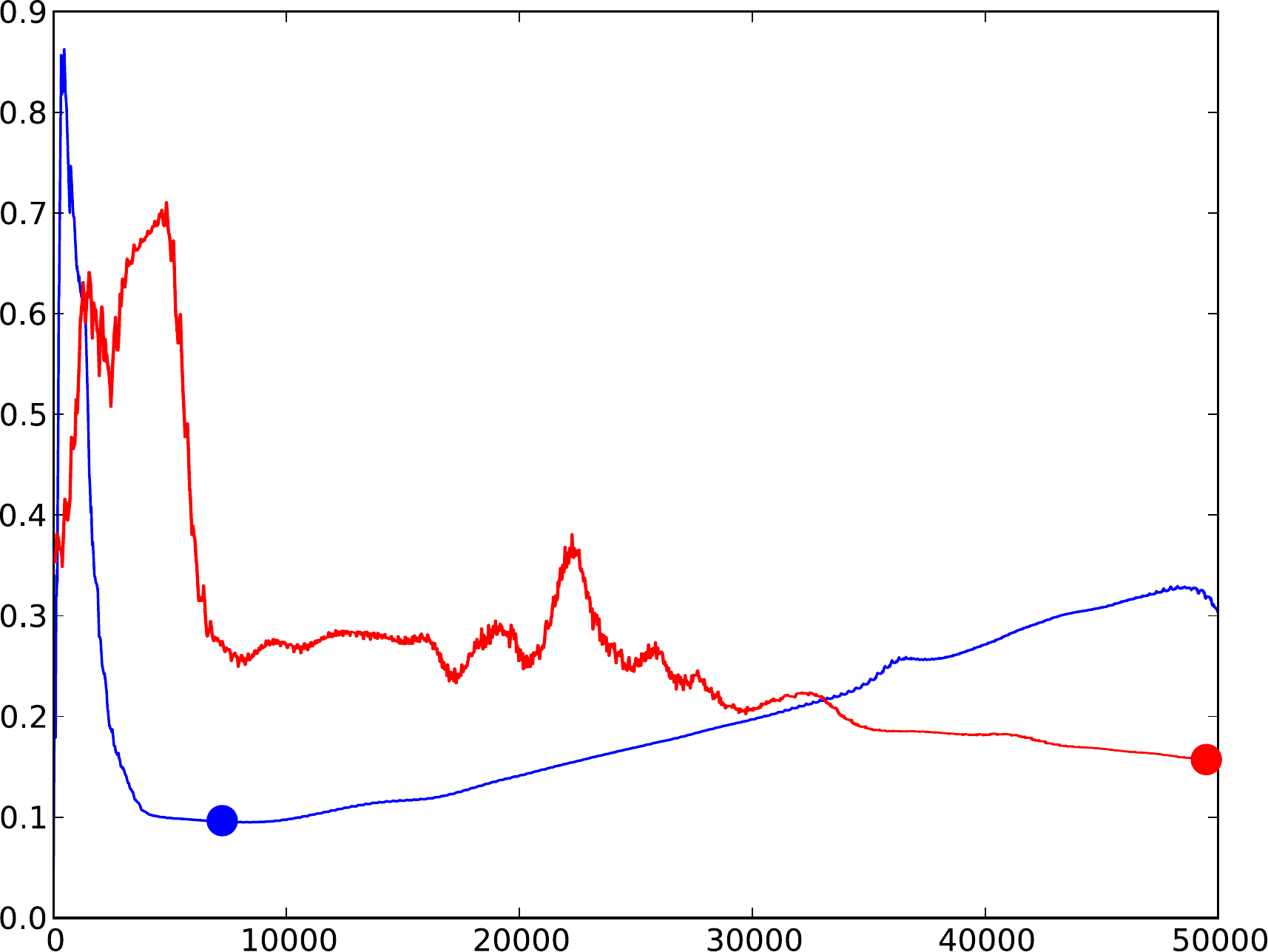}
\caption{Control performance of IDRNN (red) and DRA (blue). The $L_1$ distance between the target $t \mapsto \z_{t}$ and $t \mapsto \u_{t + \delta}$ over a siding window of size $900$ is used for this performance index. The two dots correspond to figure \ref{fig: heating tank control}.}
	\label{fig: heating tank perf}
\end{figure}

\begin{figure}[htbp]
	\centering
		\subfigure[IDRNN control.]{\includegraphics[width=.45\textwidth]{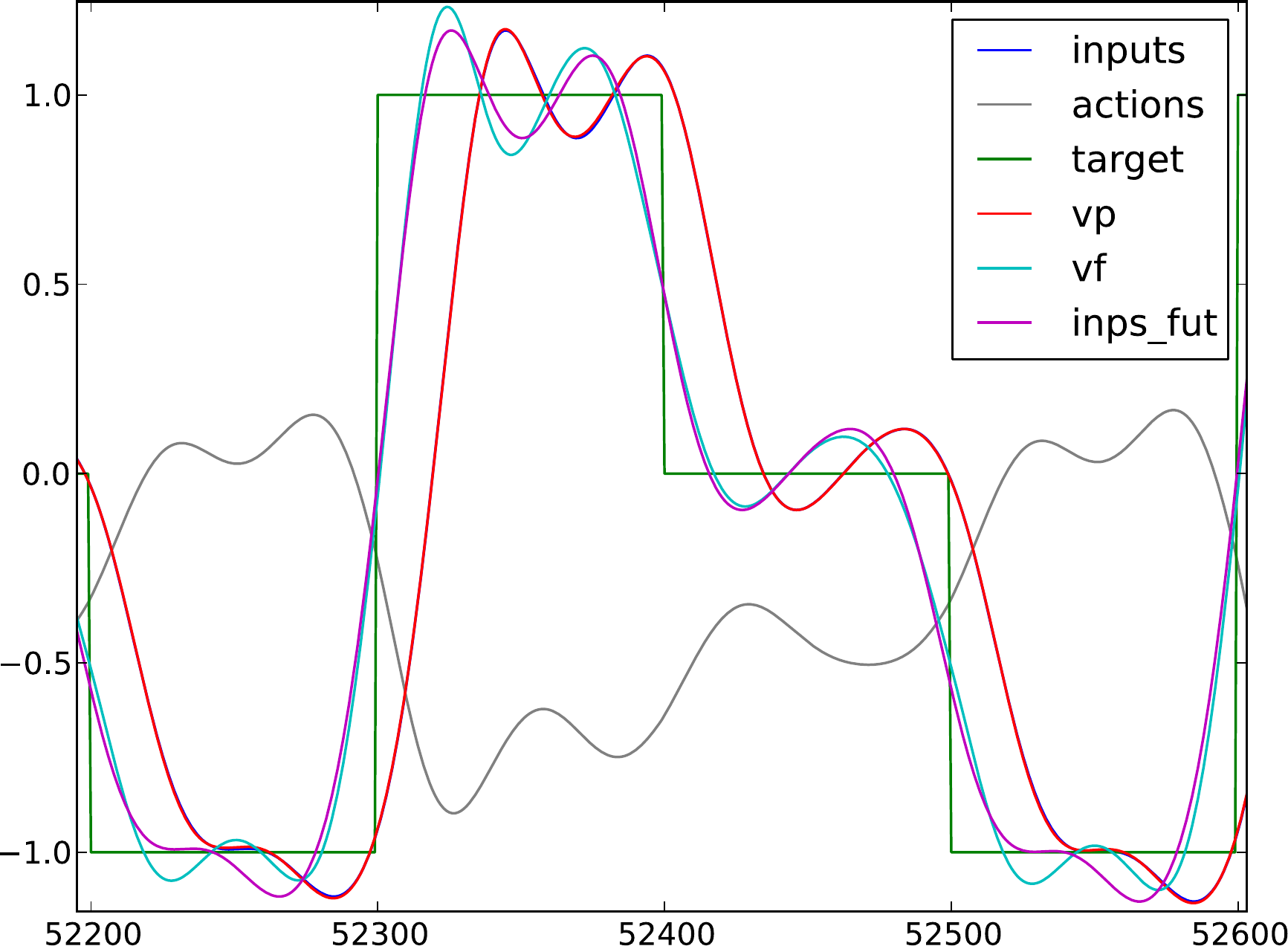} \label{fig: IDRNN traj}}\qquad
		\subfigure[DRA control.]{\includegraphics[width=.45\textwidth]{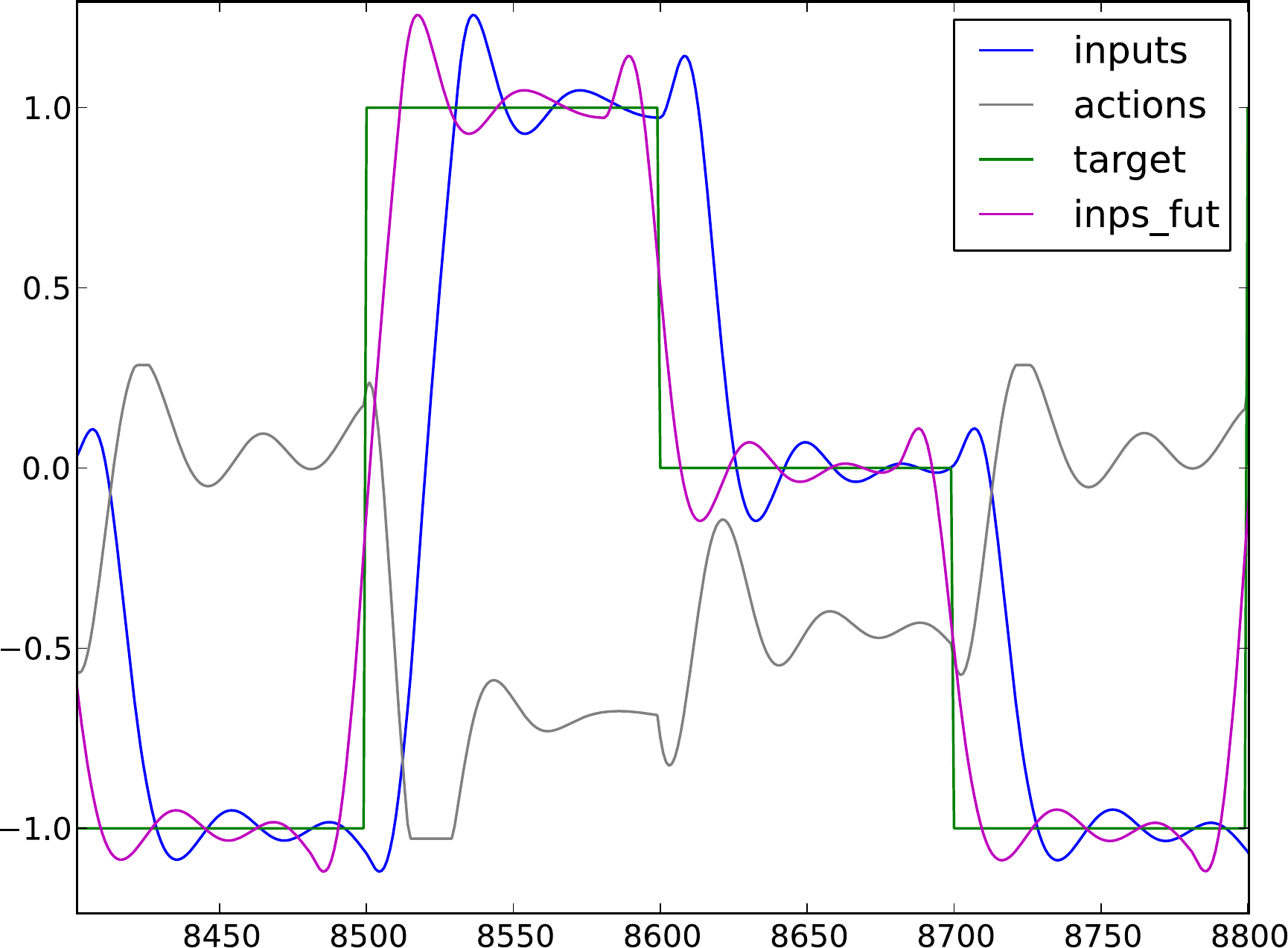} \label{fig: double res traj}}
	\caption{Results of the control of the heating tank for the IDRNN (left) and DRA (right). The IDRNN control corresponds to the red dot in figure \ref{fig: heating tank perf} and the DRA control corresponds to the blue dot in figure \ref{fig: heating tank perf}. In blue are the inputs $\u_t$ and in pink are the future inputs $\u_{t + \delta}$. In grey are the actions $\va_t$. In green is the target $\z_t$. In red is the perception $\vp$ and in cyan in the prediction of the future $\vf$.}
	\label{fig: heating tank control}
\end{figure}

I simulated the IDRNN and the DRA control for $50000$ time steps and obtained a performance displayed in figure \ref{fig: heating tank perf}. It appears that the DRA converges faster to a good control behavior, see \ref{fig: double res traj} but its performance slowly deteriorates with time. On the contrary, the IDRNN converged more slowly but its performance  improves with time. Eventually, the system behavior (see \ref{fig: IDRNN traj}) is similar to the best case for the DRA, although a bit worse. The parameters used are the same as in the previous section with the addition of $\delta = 20$, $\lambdap = \lambdaa = \lambdaf = 0.9999$, $\mup =\muf = \mua = 10^{-6}$. The reason why I take the forgetting parameters $\lambdap = \lambdaa = \lambdaf$ to be so large is because an initializing method is used and I want the network not to forget immediately the initialization. I also reset the matrix $\P_a$ to $0.002 I_d$ at the end of the initialization to slow learning down to stay close to initialization at first. The parameters for the DRA are identical to that of the IDRNN.

To summarize, the two methods seem to roughly give similar results at their optimum. However, they seem to differ in the evolution of the performance with time.

\section{Discussion}\label{sec: discussion}
{
Beyond the proof of concept supported by the previous numerical experiments, IDRNN has some notable properties which are discussed in this section.
}
\subsection{Reproducing other control schemes}
{
An important characteristic of IDRNN (and also DRA) is its universality, in the sense that it can embed any existing control schemes. Indeed, as detailed in the previous section, it is possible to initialize these networks, using traditional ESN algorithms. The procedure is, first, to record the inputs (perception) and (output) of a preexisting control scheme for a given environment; and, second, to use traditional ESN algorithms on the IDRNN architecture in order to reproduce these trajectories when exposed to a similar environment.
}

{
Sontag has proven that, in theory, recurrent neural networks could reproduce any dynamical systems \cite{sontag1997recurrent}. Thus, there exists an IDRNN, possibly with a very large number of neurons, which can emulate a given control scheme arbitrarily accurately. Naturally in applications, the number of neurons is limited and the reproduction is not necessarily accurate, even more if the recorded trajectories were not long enough (see previous section). Nonetheless, this method can be used as a precious initialization procedure in order to design incremental control schemes.
}

\subsection{Extension to reinforcment learning}
{
Although the present version of IDRNN is explicitly designed for a supervised learning framework, it possible to extend the approach to a reinforcment learning framework. {The first step is to include a reward in the present model (which the neural network will try to maximize). This can be easily done by adding a stimulus neuron exclusively excited when a reward in presented. Thus, implementing a reinforcment learning approach would simply correspond to maximizing the activity of this reward neuron instead of minimizing the distance between such a neuron and a target trajectory. In the mathematical formalism, this can be immediately implemented by replacing the target $\z_t$ by $0$ in \eqref{eq: ideomotor principles} and asking the motor learning to maximize (rather than minimize) its criterion . Thus, motor learning aims at increasing the current and future rewards.}
}

{
Besides, the network can also handle a variety of hybrid approaches corresponding for instance to a mix of supervised / reinforcment learning. More precisely, an interesting situation is to design an agent with a lot of stimuli (including a reward) all of which perceptive learning aims at predicting, while motor learning exclusively aims at increasing the reward. This would correspond to a common reinforcment learning situation where there is no target, while making sure the network can behave coherently and in context with its environment.
}

{
The IDRNN approach has, in principle, no problem handling rare or intermittent rewards. Indeed, it still produces behavior when the rewards are very sparse. The IDRNN network activity is always running and not directly influenced by the presence of a reward. Only learning is directly boosted when a reward is presented. In my opinion, this provides a notable advantage of IDRNN over DRA which can not be extended to reinforcment learning so easily because the activity of the network is directly influenced by the target/reward. Thus with rare rewards, i.e. sparse target, the DRA approach would not work.
}

{
The proper handling of distal rewards \cite{jordan1992forward} by the IDRNN is still an open question, but the machinery introduced here for prediction of the future in several time steps seems to be an appropriate way to treat this difficult problem. Nonetheless, this stands as a perspective to be investigated.
}

\subsection{Biological plausibility}
{
Of course, I do not claim the full biological plausibility for the IDRNN approach because such simple networks can never represent the incredible complexity found in neuroscience. However, I would argue that this approach is biologically more plausible than the DRA. Indeed, the latter approach uses two reservoirs with exactly the same recurrent and output weights. This clearly breaks the locality requirement of biologically plausible algorithms \cite{gerstner2002mathematical}: to modify the connection between neurons $i$ and $j$, learning rules should only include information available at neurons $i$ and $j$. On the other hand, the IDRNN approach is based on a single reservoir and there is no sharing of the weights. Thus, contrary to DRA, it is not biologically implausible by construction.
}

{
Although this paper is based on RLS implementation of the ideomotor principles (for efficiency reason), it is also possible to design of LMS implementation which is more biologically plausible. Indeed, RLS is not local since is involves computing the inverse of the global correlation matrix. LMS corresponds to a simple stochastic gradient descent of the ideomotor principles in \eqref{eq: ideomotor principles}. It leads to slower convergence times, but more robustness, than the RLS algorithm \cite{haykin2005adaptive}, but both aim at solving the same problem. The ideomotor LMS algorithm can be derived from the modified ideomotor principles \eqref{eq: RLS pple} under the same greedy assumption that $\u$ and $\vr$ do not depend on $\Wp$ and $\Wa$. It comes as the gradient descent of $H^p$ and $H^a$. Assuming that we consider the reinforcment learning case detailed above, where $\z =0$ and motor learning is a maximization, it can be written as
\begin{equation}
 \begin{array}{rl}
  \Wp_{t+1} & = \Wp_t + \eps_p (\u_t - \Wp_t.\vr_t).{\vr_t}'\\
  \Wa_{t+1} & = \Wa_t + \eps_a \Wprestr.\vr_t.{\vr_{t-1}}'\\
 \end{array}
 \label{eq: LMS algo}
\end{equation}
where $\eps_p$ and $\eps_a$ have been re-parametrized to absorb constants.
}

{
In this form, perceptive learning is local. Indeed, the modification of the connection $\{\Wp\}_{ij}$ only depends on the stimulus $\{\u\}_i$, the prediction $\{\Wp_t.\vr\}_i$ and the reservoir state $\{\vr\}_j$ which are locally available quantities between reservoir and perceptive area.
}

\begin{figure}[ht]
 \centering
 \includegraphics[width=0.4\textwidth]{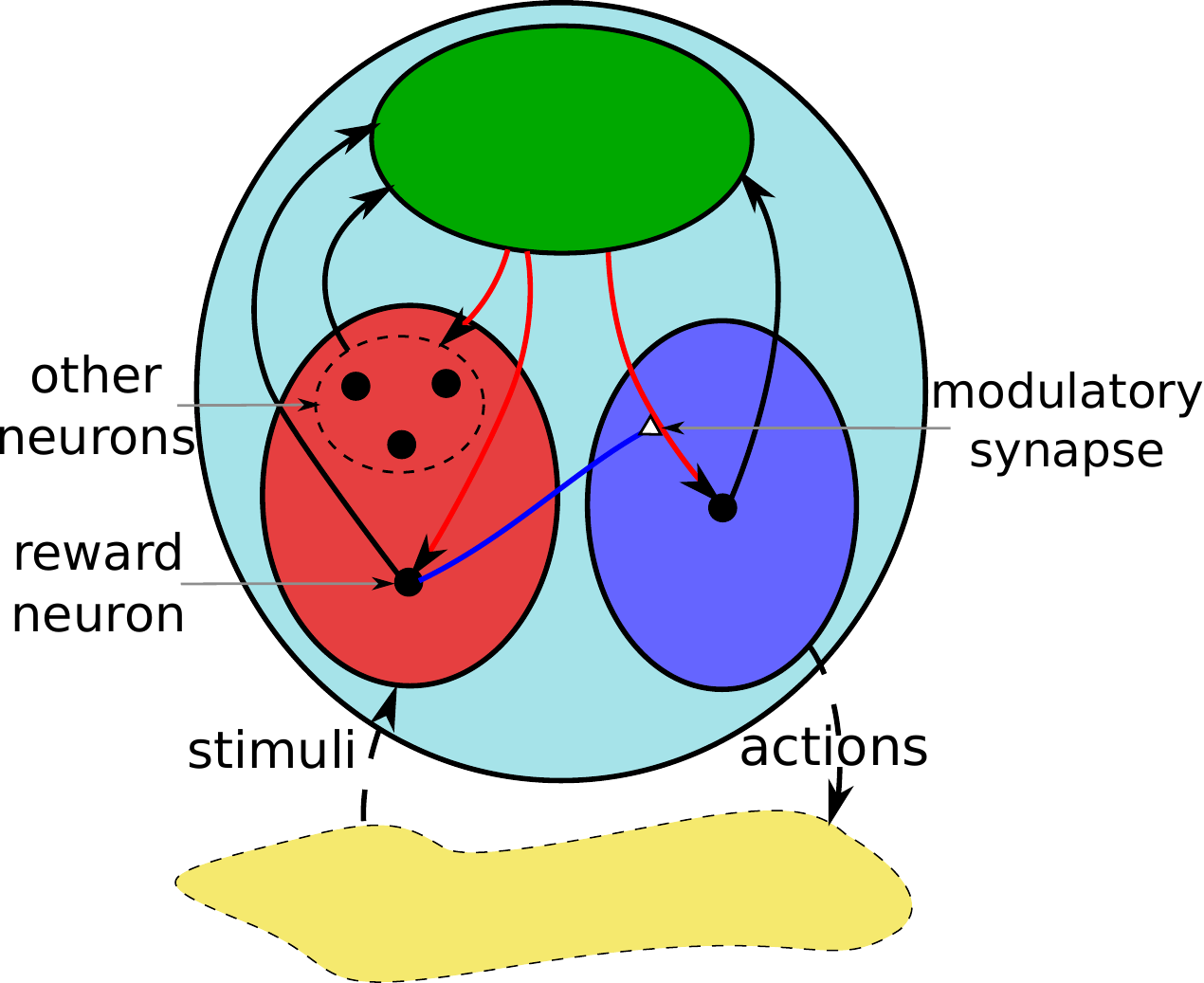}
 \caption{Diagram showing the necessary wiring of the network if learning has to be local, in the illustrative case of a single reward neuron and a single motor neuron. The perceptive area is made of a reward neuron and other perceptive neurons, all of which receive stimuli from the environment (possibly some other part of the brain). The reward neuron has to send a modulatory synapse to the motor connections. The information from the motor neuron is only used for learning and does not modify directly the dynamics of the motor neuron.}
 \label{fig: biological plausibility}
\end{figure}

{
Motor learning can also be said to be plausible, if we slightly nuance the locality requirement by the experimental fact that a few modulatory synapses can bring some additional information to distant connections as shown in figure \ref{fig: biological plausibility}. Indeed, the modification of the connection $\{\Wa\}_{ij}$ depends on the reservoir state ${\vr}_j$, but also on the $\Wprestr.\vr$ which is not available between reservoir and motor area. Thus there is a need for a modulatory synapse, involved mainly in learning, from the perceptive area to the motor connections to biologically implement this algorithm.
}

{
Interestingly, LMS and RLS can be shown to be equivalent if the reservoir has a temporal correlation matrix proportional to the identity \cite{farhang1998adaptive}. This regime has been often observed in biology, where in vivo cortical tissues are said to be in an asynchronous irregular state \cite{ecker2010decorrelated, renart2010asynchronous}. This would support the fact that such a biological LMS implementation of the ideomotor principle can be efficient, although showing this rigorously stands as a perspective.
}

{Besides, one might ask how the current scheme differs from active inference that uses explicit forward or generative models of predicted stimuli.  They key difference is the simplicity of the current scheme - that just involves optimising connection weights from the reservoir to action and perception states.  Heuristically, this can be regarded as equipping an agent with a vast repertoire of generative models and then optimising the weights to select the model with the greatest evidence (least free energy or prediction error).  The connection between the current scheme and active inference may be important from the point of view of biological implementation: there is now a literature on biological schemes for minimising prediction error using hierarchical predictive coding and Bayesian filtering schemes.  Of particular interest here is the role of reflexes in mediating action.  The current scheme can be regarded as selecting a forward model of sensations.  In active inference, these forward models also predict kinaesthetic or proprioceptive sensations.  This means that the prediction errors minimised by action can be resolved very simply - through peripheral reflex arcs \cite{adams2013predictions}.}

{
On the whole, this theory driven approach may contribute to the debate about the computational role of several parts of the brain. A rigorous link with the brain clearly stands as a long term perspective. Yet, I think it is interesting to note the ingredients this architecture needs in order to design what could considered as a simplistic embodied agent. The first ingredient is the combination of a central recurrent network and two types of read-out, as can be observed in the spinal chord \cite{butler2005comparative} with the dorsal root for perception and the ventral root for action. The second ingredient is the modulation of motor connections by reward neurons which seems to be handled in the brain by a variety of neurotransmitters \cite{seamans2008dopamine}. I believe that the study of basic vertebrate nervous system may, in the long term, benefit from such constructive approaches.
}

\section{Conclusion}
This paper defines a recurrent neural network which blindly learns to control an unknown environment. Based on a randomly connected reservoir it learns on the fly two read-outs which correspond to perception and action. These read-outs are learned according to two principles: perceptive learning corresponds to maintaining good predictions of the incoming stimuli; motor learning tries to change the dynamics of the reservoir so that the stimuli predictions match a target trajectory. Actually, the control of the environment is just a byproduct of the behavior of the neural network which only cares about its own predictions. This algorithm is closely {related to the ideomotor theory and active inference, providing an efficient computational implementation of these concepts.}
{An implementation of the proposed method to robotics may highlight the similarities with more classical approaches in this field \cite{tani1996model}.}

Several numerical simulations have established a proof of concept for this neural network. It manages to control fairly complicated dynamical systems and properly handles non-linearities and delays. {The robustness of the approach with respect to most parameters is supported by the fact that a single set of parameters (with little tuning) was used to control all environments in this paper.} 

{Several challenges can be foreseen for the future development of such algorithm. First, a extensive benchmarking of IDRNN, DRA and competitors on real world environments is needed. Second, the development of a mathematical theory explaining the power of such random networks would be useful. {Third, extending the ideomotor approach presented here to a fully connected neural network (e.g. with connections from perceptive to motor area) can be done straightforwardly, and may prove more efficient in certain cases.} Finally, building architectures with building blocks such as the network presented here could prove interesting, not only for designing even more intelligent agents, but also to shed light on possible information hierarchies which might be implemented by the brain.}

\paragraph{Aknowledgments:} 
I thank Herbert Jaeger, Michael Thon, Jochen Steil, Felix Reinhart and Benjamin Schrauwen for helpful discussions. I was funded by the European project AMARSI. 


\end{document}